%% file: Predict2011TwoColumn.tex
\documentclass[10pt,journal,final,finalsubmission,twocolumn]{IEEEtran}

\pdfoutput=1

\input{header.tex}
\input{input.tex}

\begin{document}

\title{Grassmannian Predictive Coding for Limited Feedback in Multiple Antenna Wireless Systems}

\author{Takao~Inoue,~\IEEEmembership{Member,~IEEE} and
        Robert~W.~Heath,~Jr.,~\IEEEmembership{Fellow,~IEEE}%
\thanks{This material is based in part upon work supported by the National Science Foundation under grant CCF-830615. This work has appeared in part in the 2011 IEEE Int. Conf. on Acoustics, Speech and Signal Process.}
\thanks{Takao Inoue is with National Instruments, 11500 N. Mopac Expwy, Austin, TX 78759 USA. Email: takao@ieee.org.}%
\thanks{Robert W. Heath, Jr. is with The University of Texas at Austin, Department of Electrical and Computer Engineering, Wireless Networking and Communication Group, 1 University Station C0803, Austin, TX, 78712-0240 USA. Email: rheath@ece.utexas.edu.}
}


\maketitle

\begin{abstract}
\boldmath
Limited feedback is a paradigm for the feedback of channel state information in wireless systems. In multiple antenna wireless systems, limited feedback usually entails quantizing a source that lives on the Grassmann manifold. Most work on limited feedback beamforming considered single-shot quantization. In wireless systems, however, the channel is temporally correlated, which can be used to reduce feedback requirements. Unfortunately, conventional predictive quantization does not incorporate the non-Euclidean structure of the Grassmann manifold. In this paper, we propose a Grassmannian predictive coding algorithm where the differential geometric structure of the Grassmann manifold is used to formulate a predictive vector quantization encoder and decoder. We analyze the quantization error and derive bounds on the distortion attained by the proposed algorithm. We apply the algorithm to a multiuser multiple-input multiple-output wireless system and show that it improves the achievable sum rate as the temporal correlation of the channel increases.
\end{abstract}


\IEEEpeerreviewmaketitle

\section{Introduction}
Multiple antenna wireless communication systems can improve throughput and reliability when channel state information (CSI) is known at the transmitter. Limited feedback is a flexible approach for providing quantized channel state information from the receiver to the transmitter. Most prior work on limited feedback use one-shot feedback that makes an instantaneous channel measurement and sends back the quantized CSI without memory. In a mobile environment, however, the channel exhibits coherence over time that may be exploited to improve the resolution of the quantized CSI at the transmitter.

Predictive vector quantization (PVQ) is a class of memory based coding techniques used in applications such as speech, image, and video processing \cite{Haoui1984,Hang1985,Gersho1991,Khalil2001}. In PVQ, the error signal between the current observed vector and the predicted vector based on past observations is quantized. When the observed data to be encoded are correlated, usually in time or space, quantizing the error signal leads to lower distortion compared with memoryless vector quantization~\cite{Gersho1991}. The effectiveness of PVQ rests on the correlation exhibited by the data, the prediction function, and the quantization technique employed. Due to temporal correlation in the propagation channel, it is natural to consider predictive coding approach for encoding CSI in temporally correlated channels. Classical PVQ has been applied for signals in linear vector space where the usual difference, addition, and prediction are well understood. Unfortunately, in multiple antenna limited feedback beamforming in wireless communication, the CSI to be encoded lives often on the Grassmann manifold. The Grassmann manifold, denoted $\cG_{n,p}$, is the set of $p$-dimensional subspaces of $n$-dimensional Euclidean space. Because it is a nonlinear manifold, extending classical PVQ is challenging since the usual linear operations, not to mention important functions like prediction, are not well defined. 


Motivated by applications in multiple-input multiple-output (MIMO) wireless communication, there has been research in analyzing~\cite{Barg2002}, quantizing~\cite{Dai2005,Mondal2007,Ashikhmin2007}, and coding~\cite{Zheng2002,Kammoun2003,Cipriano2005} on the Grassmann manifold driven in part by applications to commercial wireless systems~\cite{80216e,3GPP}. Prior work exists for designing suitable memoryless quantization codebooks such as Grassmannian line packing~\cite{Narula1998}, vector quantization~\cite{Roh2006}, Grassmannian frames~\cite{Tropp2005}, and Kerdock codebooks~\cite{Inoue2009b} (e.g., also see the references in \cite{Love2008}). Several techniques have been previously proposed to exploit the temporal correlation of the propagation channel~\cite{Huang2006,Huang2009,Mondal2004b,Samanta2005,Kim2008,Raghavan2007,Heath2008,Liu2006a,Kim2011a}. In \cite{Huang2006,Huang2009}, modeling the feedback state transitions allow the net feedback rate to be reduced. The resolution, however, is fixed by the codebook size. To improve the quantization error, an adaptive codebook approach was proposed that can adapt to a given channel distribution \cite{Mondal2004b}. Additional feedback overhead to retrain or synchronize the pre-computed codebooks may be needed when the channel distribution changes. Alternatively, a hierarchical codebook strategy uses two codebooks, coarse and fine, for layered feedback in temporally correlated channel~\cite{Samanta2005,Kim2008}. A codeword describing the coarse encoding region is updated infrequently and a finer local codebook is used for frequent feedback. A more flexible approach is to use a progressive refinement strategy in which rotation and scaling are applied to structured codebook so as to provide high resolution feedback~\cite{Raghavan2007,Heath2008}. An approach related to our paper is the complex Householder transform based PVQ-like technique for correlated normalized channel vectors in multiple-input single-output communication systems~\cite{Liu2006a}. The current vector channel is decomposed into previous vector channel and weighted sum of orthogonal subspaces to represent the temporal variation. While the algorithm is presented in the form of PVQ, the actual operation is successive decomposition and projection using the complex Householder transform with unit delay which was shown to be optimal for the specific application. A differential feedback approach using a rotation codebook has been proposed for spatial multiplexing system \cite{Kim2011a}. They require long term correlation statistics to design suitable codebook and exploit the structure of the Riemannian manifold. Unfortunately, the codebooks are specific to the given long term statistics and may become outdated. The Grassmannian predictive coding technique proposed in this paper was presented in part in \cite{Inoue2011}. We proposed the Grassmannian predictive coding algorithm applied to multiuser MIMO system. We did not, however, provide the details of derivation nor consider an efficient codebook representation and a distortion analysis. 

In this paper, we propose a predictive coding algorithm for correlated data on the Grassmann manifold, which we call the Grassmannian predictive coding (GPC) algorithm. The GPC algorithm is derived using the intrinsic geometry of the manifold and corresponding mathematical operations that respect the curved manifold structure. The main contributions of this paper are as follows.
\begin{itemize}
  \item {\it Grassmannian predictive coding algorithm}: We propose a framework for predictive coding on the Grassmann manifold. The key idea of our approach is to use the {\it tangent vector} to establish the notion of a difference between points on the manifold. The proposed prediction function uses {\it parallel transport} as a one step prediction. The prediction step uses the immediate past difference; formulating higher order prediction function remains for future work. The concepts of tangent vector and parallel transport have been used in~\cite{Edelman1998} for optimization problems, but have not been exploited to develop a predictive coding concept. 

  \item {\it Efficient codebook structure}: A design of tangent space codebook using Lloyd algorithm is proposed. The codebook lives in the tangent space of $\bbC^{\Nt}$ with magnitude dependent on the correlation exhibited by the channel and prediction function. An efficient codebook storage strategy is proposed exploiting the direction and magnitude decomposition of the tangent space vector.

  \item {\it Distortion bounds:} Based on a geometric interpretation of our GPC algorithm, a simple model of the quantization region is obtained. Using metric volume computations on the Grassmann manifold~\cite{Mondal2007}, lower and upper bounds on the quantization error are derived. We compare the obtained bounds with distortion obtained in simulations. Furthermore, we show that the distortion for the proposed GPC algorithm is lower than the lower bound of memoryless quantizer distortion for a given codebook size.

  \item {\it Application to limited feedback multiuser MIMO systems}: We apply the GPC algorithm for limited feedback zero-forcing multiuser MIMO systems with multiple transmit antennas and a single receive antenna at each mobile terminal~\cite{Jindal2006}. We show that the proposed GPC algorithm provides substantial sum rate improvement over memoryless random codebook technique with same feedback rate~\cite{Jindal2006}. The sum rate improvement, however, depends on the channel correlation. When the channel is highly correlated, the proposed GPC algorithm is shown to provide sum rates close to a system with perfect CSI at the transmitter, i.e., infinite feedback.
\end{itemize}

{\it Notation:} We use lower case bold letters, e.g., ${\bf v}$, to denote vectors and upper case bold letters, e.g., ${\bf H}$, to denote matrices. A 2-norm is denoted by $\| \cdot \|$ and a normalized vector is denoted by $\vec{\bf v} = {\bf v}/\|{\bf v}\|$. The $n \times n$ identity matrix is denoted by $\bI_n$. The space of integers and complex numbers are denoted by $\bbN$ and $\bbC$, respectively, with an appropriate superscript to denote the dimension of the respective spaces. We use $^T$, $^*$, and $^\dagger$ to denote the transposition, Hermitian transpose, and pseudo inverse, respectively. The $n$-th column entry of a matrix $\bA$ is denoted by $[\bA]_{:,n}$. The expectation is denoted $\bbE[\cdot]$. 

\section{System Model}\label{sec:system}\label{sec:system:mumimo}
In this paper we apply GPC algorithm to limited feedback multiuser MIMO communication. It can also be applied to single user MIMO and to multi-cell MIMO systems.  Multiple user MIMO is a challenging application of limited feedback as it requires high resolution quantization \cite{Jindal2006} and is known to be sensitive to channel variations \cite{Huang2007c}. We consider a multiuser limited feedback system with $\Nt$ transmit antennas at the base station and $U \leq \Nt$ mobile users each equipped with a single receive antenna. To isolate the impact of using predictive coding for limited feedback, we assume that $U$ users are scheduled a priori from possibly large number of user pool; we do not consider scheduling or the effects of multiuser diversity in this paper. Let $s_u[k]$, $\bv_u[k]$, and $\bh_u[k]$ be the complex transmit symbol, $\Nt \times 1$ unit norm beamforming vector, and $\Nt \times 1$ channel vector for $u$-th user at time index $k$, respectively. We assume that the transmit vector $\bs = [s_1[k] \cdots s_U[k]]^T$ satisfies the total transmit power constraint $\bbE[\|\bs\|^2] \leq P$. Then, the input-output relationship for $u$-th user may be written as
\begin{equation}
  y_u[k] = \bh_u^*[k] \bv_u[k] s_u[k] + \bh_u^*[k] \sum_{n=1,n \neq u}^{U} \bv_n[k] s_n[k] + n_u[k]
  \label{eq:mumimo_io}
\end{equation}
where $n_u$ is an independent identically distributed (i.i.d.) zero mean complex Gaussian noise with unit variance at user $u$. The first term in \eqref{eq:mumimo_io} is the desired signal for $u$-th user while the second summation term is the interference signal. The signal to interference plus noise ratio (SINR) for the $u$-th user can be written as
\begin{equation}
  \text{SINR}_u = \frac{\frac{P}{U} |\bh_u^* \bv_u|^2}{1+\sum_{n \neq u} \frac{P}{\Nt} |\bh_u^* \bv_n|^2}.
  \label{eq:mumimo_sinr}
\end{equation}
If the transmit signal $s_u$ is assumed to be Gaussian, the achievable rate for user $u$ is given by
\begin{equation}
  \cR_u = \log_2 (1+ \text{SINR}_u)
  \label{eq:mumimo_rate}
\end{equation}
and the sum rate as ${\bf \cR} = \sum_{u=1}^U \cR_u$. 

 The SINR expression \eqref{eq:mumimo_sinr} shows that the amount of interference depends on the design of the beamforming vectors. Zero forcing uses beamforming vectors such that they are orthogonal to other user's channel vectors, i.e., $\bh_u[k] \bv_u[k] = 0$ for $n \neq u$, to null the inter user interference \cite{Caire2003}. Let $\bH = [\bh_1 \cdots \bh_u]^*$ be the $U \times \Nt$ composite channel matrix. With perfect CSI, the interference can be completely eliminated by choosing the unit norm beamforming vector as the normalized columns of pseudo inverse composite channel matrix, i.e., $\bv_u = [\bH^\dagger]_{:,u}/\|[\bH^\dagger]_{:,u}\|$. Zero forcing creates $U$ interference free parallel channels providing nearly linear throughput increase as a function of number of users but with some power loss due to normalization~\cite{Peel2005}. 

In limited feedback multiuser MIMO systems, quantized CSI is fed back to the transmitter from each user~\cite{Jindal2006,Huang2007c}. Assuming that a perfect channel estimate $\bh_u$ is obtained, we consider the quantization of the channel direction $\bg_u = \bh_u / \|\bh_u\|$ and assume that the scalar channel gain is known perfectly~\cite{Huang2007c}. We assume that the channel gain is dependent on the longer term statistics that varies much slower than the channel direction. Since the channel gain is a real valued quantity that is easier to feedback, we assume that the channel gain is known perfectly at the transmitter and consider the effects of the channel shape quantization only \cite{Huang2007c}. 
In this regime, the SINR can be rewritten as
\begin{equation}
  \text{SINR}_u = \frac{\frac{P}{\Nt} \|\bh_u \|^2 |\bg_u^* \bv_u|^2}{1+\sum_{n \neq u} \frac{P}{\Nt} \|\bh_u\|^2 |\bg_u^* \bv_n|^2}.
  \label{eq:mumimo_shape_sinr}
\end{equation}
We make two observation from \eqref{eq:mumimo_shape_sinr}. First, if the channel vector $\bh_u$ is an i.i.d. vector distributed according to $\cC\cN(0,1)$, $\bg_u$ is isotropically distributed on the $\Nt$-dimensional hyper-sphere. Second, due to the absolute value around $\bg_u^* \bv_u$, the SINR is independent of arbitrary unitary rotations of the channel direction. That is, $|\bg_u^* \bv_u|^2 = |e^{j\theta} \bg_u^* \bv_u|^2$ for $\theta \in (0,2\pi]$. Therefore, we may identify the space of channel shape as the Grassmannian manifold. Thus, the problem of transmit beamformer design is to feedback channel shapes on the Grassmann manifold from each user $u$, and use the collected channel shape information at the transmitter to design the beamforming vectors by zero forcing. 

In conventional codebook based limited feedback multiuser MIMO systems, each user has a normalized channel vector codebook of size $N_{\text{RC}}$ which is shared with the transmitter~\cite{Jindal2006,Huang2007c}. The transmitter maintains $U$ tables of size $N_{\text{RC}}$ codebooks. Each user selects the codeword with minimum chordal distance from the normalized channel vector estimate. The index of the selected codeword using $\log_2(N_{\text{RC}})$ bits is fed back to the transmitter. The transmitter collects the decoded channel vectors $\hat{\bh}_u$ for each user $u$ to form the composite channel matrix $\hat{\bH} = [\hat{\bh}_u \cdots \hat{\bh}_u]^*$. The beamforming vectors are computed as $\hat{\bv}_u = [\hat{\bH}^\dagger]_{:,u}/\|[\hat{\bH}^\dagger]_{:,u}\|$. Using a random codebook, it was shown in \cite{Jindal2006} that sum rate performance becomes interference limited as signal to noise ratio (SNR) increases and that codebook size needs to be increased linearly as a function of SNR, in dB, to maintain multiplexing gain. Herein, lies the practical limitation of the conventional codebook approach: the codebook size that approaches the achievable sum rate becomes impractical even for moderate SNR. The proposed GPC algorithm overcomes this problem. 


\section{Grassmann Manifold: Preliminaries}\label{sec:prelim}
The geometric and linear algebraic properties of the Grassmann manifold will be fundamental in derivation of our proposed algorithm. In this section we review key definitions, properties, and mathematical tools pertaining to designing algorithms for the Grassmann manifold. Then we propose a predictor on the Grassmann manifold built from the tangent vector, mapping from the tangent onto the manifold, and parallel transport. 

Let $\cU_n = \{\bX \in \bbC^{n \times n}: \bX^* \bX = \bI_n\}$ be the unitary group formed by $n \times n$ unitary matrices. For $p<n$, the Grassmann manifold, $\cG_{n,p}$, is the set of subspaces spanned by the columns of the quotient group $\cU_n / \cU_{n-p}$. It may also be identified as the quotient space of the unitary group, $\cU_n / (\cU_{n-p} \times \cU_p)$. A point $\bX \in \cG_{n,p}$ may be considered as an equivalence class, i.e., $[\bX] \bydef \{\bX\bU_p: \bU_p \in \cU_p\}$. For notational brevity, we denote $\bX \in \cG_{n,p}$ to mean the equivalence class of matrices whose columns span the same $p$-dimensional subspace. For numerical computation, we understand $\bX \in \cG_{n,p}$ to be one representative of the equivalence class. The Grassmann manifold is a smooth topological manifold with a locally Euclidean property and smooth tangent space structure~\cite{Lee2003}, both of which will be essential in the derivation of the proposed algorithm. In this paper, we consider the Grassmann manifold $\cG_{n,1}$; the general case of $p >1$ is a topic of future work.

Let the inner product of $\bx,\ \by \in \cG_{n,1}$ be denoted by $\rho = \bx^* \by$. Let $\theta = \cos^{-1}(|\rho|)$ be the subspace angle between $\bx$ and $\ \by$~\cite{Golub1996}. The chordal distance metric for $\cG_{n,1}$ is given by~\cite{Edelman1998,Love2003}
\begin{eqnarray}
  d(\bx,\by) & = & \sqrt{1-|\rho|^2} \nonumber \\
    & = & | \sin \theta |.
  \label{eq:chordal_dist}
\end{eqnarray}
For notational brevity, we use $d$ without the arguments when there is no confusion. Unlike the arc length, given by $| \theta |$, the chordal distance is differentiable everywhere and provides a close approximation of the arc length when the points are close~\cite{Conway1996}. 

Using the chordal distance metric, we define the correlation of two sequences $\{\bx[k]\}_{k \in \bbN}, \{\by[i]\}_{i \in \bbN} \in \cG_{n,1}$ by $\zeta_{\bx,\by}[n] = \bbE_k[d(\bx[k],\by[k+n])]$ which can be interpreted as the mean chordal distance between two sequences on the Grassmann manifold. 

\begin{figure}[!htbp]
\centering
\includegraphics[width=0.9\columnwidth]{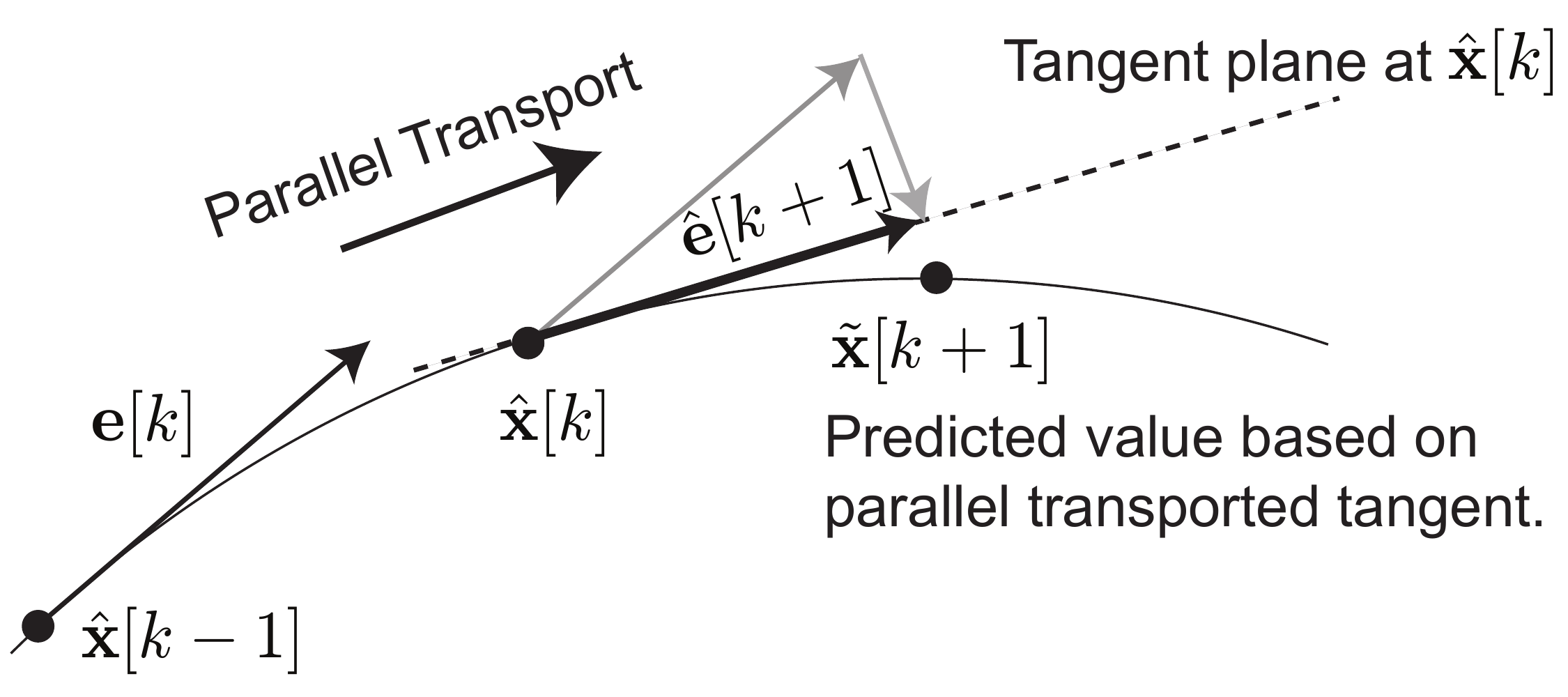}
\caption{Conceptual illustration of the tangent vector and the parallel transport on the Grassmann manifold.}
\label{fig:PT}
\end{figure}

Based on the smooth manifold structure of the Grassmann manifold, it is possible to relate two points $\bx[k],\bx[k+1] \in \cG_{n,1}$ by considering the tangent vector emanating from $\bx[k]$ to $\bx[k+1]$. \figref{fig:PT} illustrates the concept. The tangent has been used successfully in the development of Newton and conjugate gradient algorithms with orthogonality constraints~\cite{Mahony1994,Edelman1998,Manton2001a,Manton2001b}. We utilize the tangent relationship for its computational benefits and geometric insight to the problem. 

\begin{lemma}[Tangent]
  If $\bx[k],\ \bx[k+1] \in \cG_{n,1}$, then the tangent vector emanating from $\bx[k]$ to $\bx[k+1]$ is
  \begin{eqnarray}
    \bee & = & \tan^{-1} \left( \frac{d}{|\rho|} \right) \frac{\bx[k+1] / \rho - \bx[k]}{\|\bx[k+1] / \rho - \bx[k]\|}
    \label{eq:tangent}
  \end{eqnarray}
  such that $\|\bee\| = \tan^{-1}(d/|\rho|)$ is the arc length between $\bx[k]$ and $\bx[k+1]$ and $$\vec{\bee} = \frac{\bx[k+1]/\rho - \bx[k]}{d/|\rho|}$$ is the unit tangent direction vector.
\label{lemma:tangent}
\end{lemma}

\begin{proof}
See Appendix~\ref{app:lemma:tangent}.
\end{proof}

Lemma \ref{lemma:tangent} provides a compact formula for the tangent vector relating points $\bx[k]$ and $\bx[k+1]$ on $\cG_{n,1}$. For notational brevity, we denote $\bee = L(\bx[k], \bx[k+1])$. The tangent vector can be interpreted as a length preserving unwrapping of the arc between $\bx[k]$ and $\bx[k+1]$ onto the tangent space at $\bx[k]$. Furthermore, it is conveniently expressed as the product of a magnitude component and the normalized directional component. The decomposition will be exploited in codebook design for efficient storage. 

The tangent vector describes the shortest distance path along the arc from $\bx[k]$ to $\bx[k+1]$, called the {\it geodesic}~\cite{Edelman1998}. The geodesic can be parameterized by a single parameter $t \in [0,1]$ using the tangent vector as the next lemma shows. 

\begin{lemma}[Geodesic]
  If $\bx[k],\ \bx[k+1] \in \cG_{n,1}$, $\bee$, $\| \bee\|$, and $\vec{\bee}$ are the tangent vector emanating from $\bx[k]$ to $\bx[k+1]$, the norm of the tangent vector, and the normalized tangent vector, respectively, then the geodesic path between $\bx[k]$ and $\bx[k+1]$ is 
  \begin{eqnarray}
    G(\bx[k], \bee, t) & = & \bx[k] \cos(\|\bee\| t) + \vec{\bee} \sin(\|\bee\| t)
  \end{eqnarray}
  for $t \in [0,1]$ such that $G(\bx[k],\bee,0) = \bx[k]$ and $G(\bx[k],\bee,1) = \bx[k+1]$.
  \label{lemma:geodesic}
\end{lemma}

\begin{proof}
See Appendix~\ref{app:lemma:geodesic}.
\end{proof}

Lemma \ref{lemma:geodesic} provides a convenient formula to relate points between $\bx[k]$ and $\bx[k+1]$ in terms of the tangent vector and the step size $t$. 
To introduce the notion of prediction, we use the tangent vector with respect to $\bx[k+1]$ such that it extends the geodesic path from $\bx[k]$ and $\bx[k+1]$. The translation of the tangent vector along the Grassmann manifold is accomplished by the {\it parallel transport}. 

\begin{lemma}[Parallel Transport]
  Let $\bx[k],\ \bx[k+1] \in \cG_{n,1}$ and $\bee$ be the tangent vector emanating from $\bx[k]$ to $\bx[k+1]$. Then, the parallel transported tangent vector emanating from $\bx[k+1]$ along the geodesic direction $\bee$ is
  \begin{eqnarray}
    \hat{\bee} = \tan^{-1} \left( \frac{d}{|\rho|} \right) \frac{\bx[k+1] \rho^* - \bx[k]}{d}.
    \label{eq:new_pt}
  \end{eqnarray}
  \label{lemma:pt}
\end{lemma}

\begin{proof}
See Appendix~\ref{app:lemma:pt}.
\end{proof}

Note that the general expression in \cite{Edelman1998} involves singular value decomposition (SVD) which is typically expensive for implementation. A compact form without an SVD on $\cG_{n,1}$ has not appeared in the literature before to the best of our knowledge. Thus Lemma \ref{lemma:pt} provides a convenient expression for transporting the {\it base} of the tangent vector from $\bx[k]$ to $\bx[k+1]$. It can be interpreted as transforming the tangent vector onto another tangent space connected by the geodesic.

Using the concepts of the tangent vector, geodesic, and parallel transport, we propose a one step prediction for $\cG_{n,1}$. 
\begin{definition}[One Step Grassmannian Prediction]
  Let $\bx[k],\ \bx[k-1] \in \cG_{n,1}$. The one step predicted vector $\tilde{\bx} \in \cG_{n,1}$ along the geodesic direction from $\bx[k-1]$ to $\bx[k]$ is 
  \begin{equation}
    \tilde{\bx}[k+1] = |\rho| \bx[k] + \rho^* \bx[k] - \bx[k-1] 
  \end{equation}
  such that $d(\bx[k], \tilde{\bx}[k+1]) = d(\bx[k-1],\bx[k])$.
  \label{def:prediction}
\end{definition}
See Appendix~\ref{app:def:prediction} for a detailed derivation. It is surprising that the predicted vector $\tilde{\bx}[k+1]$ can be computed by the knowledge of $\bx[k-1]$ and $\bx[k]$ using linear operations and the result remains on the Grassmann manifold. This simplification only happens for the case of taking a full step using $t=1$. It is also possible to consider smaller steps $t < 1$ as well as adaptive step sizes, but we defer this to future work. 

\section{Grassmannian Predictive Coding}\label{sec:PC}
In this section, we describe the proposed GPC algorithm. First, a general overview of the algorithm is provided. Second, the codebook design for encoding the error tangent vector is described. Finally, strategies for initialization are considered. 

\subsection{GPC Algorithm}\label{sec:PC:architecture}
\begin{figure}[!htbp]
\centering
\includegraphics[width=0.9\columnwidth]{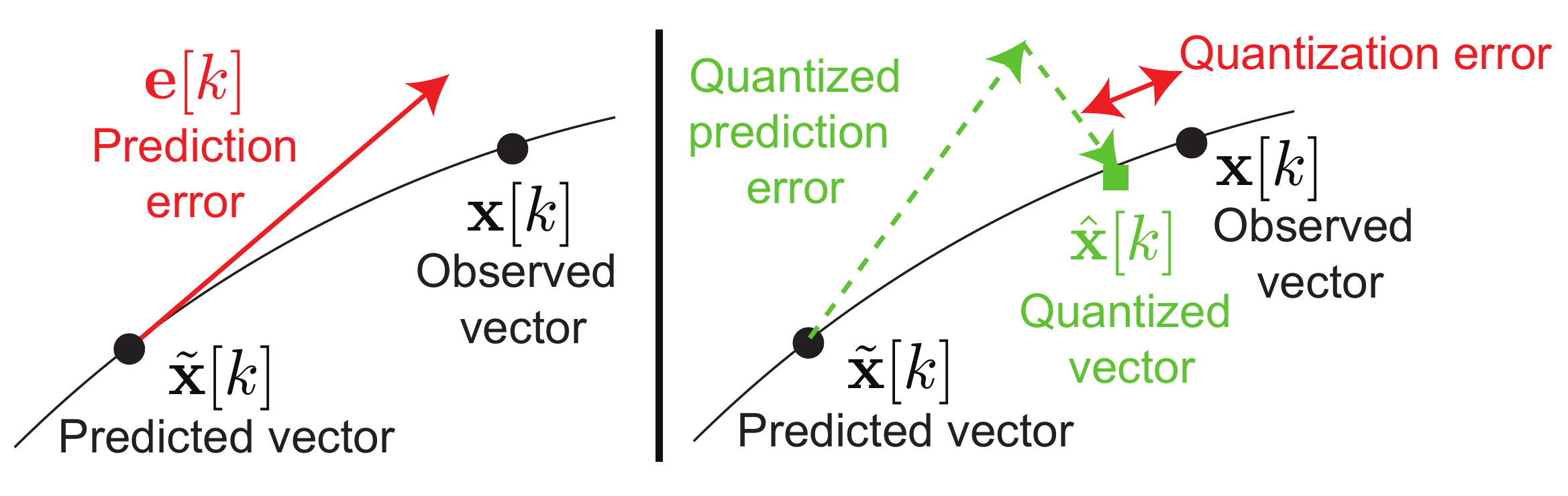}
\caption{Conceptual illustration of obtaining prediction error (left) and quantizing the prediction error to obtain the estimated vector (right).}
\label{fig:GPC}
\end{figure}

Let $\{\bx[k]\}_{k \in \bbN} \in \cG_{n,1}$ be a correlated input sequence with time index $k$. The general operation of the proposed GPC algorithm closely follows that of the conventional predictive vector quantization technique~\cite{Gersho1991}. Linear operations such as difference, quantization, addition, and prediction are replaced by equivalent operators on Grassmann manifold using the concepts derived in Section \ref{sec:prelim}. The main idea of predictive coding is to quantize the error $\bee[k]$ between the predicted vector $\tilde{\bx}[k]$ and the current observed vector $\bx[k]$. The figure on the left hand side of \figref{fig:GPC} illustrates this graphically. Then, the quantized error is applied to predicted vector to construct the state $\hat{\bx}[k]$ of the current observed vector. The figure on the right of \figref{fig:GPC} illustrates this graphically. The current and previous estimated vectors, $\hat{\bx}[k]$ and $\hat{\bx}[k-1]$, are used to compute the predict vector $\tilde{\bx}[k+1]$ as it was shown in Section \ref{sec:prelim} and  \figref{fig:PT}. Since both the encoder and decoder uses estimated vectors for prediction, they both obtain the same predicted vectors. This is in contrast to quantizing $\bx[k]$ directly in the conventional one-shot approach \cite{Love2008}. By exploiting memory, predictive vector quantization offer higher resolution for a given number of bits.

\begin{figure}[!htbp]
\centering
\includegraphics[width=0.9\columnwidth]{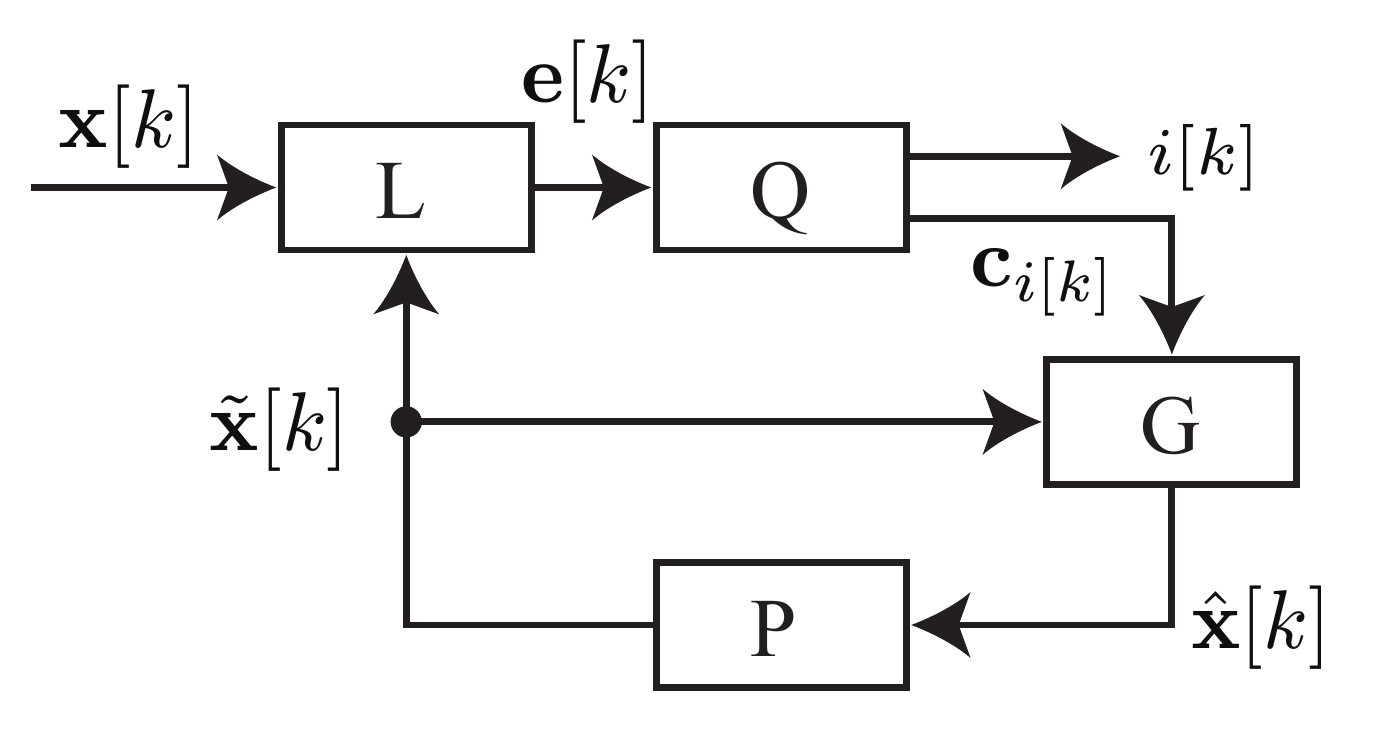}
\caption{Block diagram of predictive encoder on the Grassmann manifold.}
\label{fig:PVQ_Encoder}
\end{figure}

\figref{fig:PVQ_Encoder} illustrates the proposed GPC encoder; the pseudo code is provided in \algref{alg:GPCe}. At time $k$, an error tangent vector is computed from the predicted vector $\tilde{\bx}[k]$ to the current observed vector $\bx[k]$. Using \eqref{eq:tangent}, the error tangent vector emanating from $\tilde{\bx}[k]$ to $\bx[k]$ is computed as
\begin{equation}
  \bee[k] = \tan^{-1} \left( \frac{d}{|\rho|} \right) \frac{\bx[k] / \rho - \tilde{\bx}[k]}{\|\bx[k] / \rho - \tilde{\bx}[k]\|}
  \label{eq:tan_k}
\end{equation}
where $\rho = \tilde{\bx}^*[k] \bx[k]$ and $d = \sqrt{1-|\rho|^2}$. 

If $\cC = \{\bc_i \}_{i=1}^{N_C}$ is the size $N_C=2^b$ codebook of error tangent vectors, the index of the quantized error tangent vector is obtained by 
 \begin{eqnarray}
  i[k] & = & \argmin_{i \in \{1,2,\dots,N_C\}} d( G(\tilde{\bx}[k], \bc_i, 1), \bx[k]).
  \label{eq:general_code_select}
\end{eqnarray}
The corresponding codeword is $\bc_{i[k]}$. The codeword that yields the geodesic map with shortest distance to the observed vector $\bx[k]$ is selected. For notational brevity, we denote the quantization step by $Q:\bbC^n \to \bbN$ that takes the error tangent vector and outputs the codeword index, i.e., $i[k] = Q(\bee[k])$. The design of the codebook and efficient representation of the codebook for implementation will be described in \ref{sec:PC:codebook}. 

Continuing at the encoder, the estimated vector becomes
\begin{equation}
  \hat{\bx}[k] = G(\tilde{\bx}[k], \bc_{i[k]}, 1).
  \label{eq:estimate}
\end{equation}
Finally, the prediction using Definition \ref{def:prediction} is performed using two previous estimates
\begin{equation}
  \tilde{\bx}[k+1] = |\rho| \hat{\bx}[k] + \rho^* \hat{\bx}[k] - \hat{\bx}[k-1]
  \label{eq:predict}
\end{equation}
where $\rho = \hat{\bx}[k]^* \hat{\bx}[k-1]$. For notational brevity, we denote the prediction operation by a map $P:\cG_{n,1} \times \cG_{n,1} \to \cG_{n,1}$ which takes current and previous state vectors and outputs the predicted vector, i.e., $\tilde{\bx}[k+1] = P(\hat{\bx}[k-1],\hat{\bx}[k])$. The predicted vector is used in the next step to compute the error tangent vector. The encoding procedure is repeated for each time $k+1, k+2, \dots$.

\begin{figure}[!htbp]
\centering
\includegraphics[width=0.9\columnwidth]{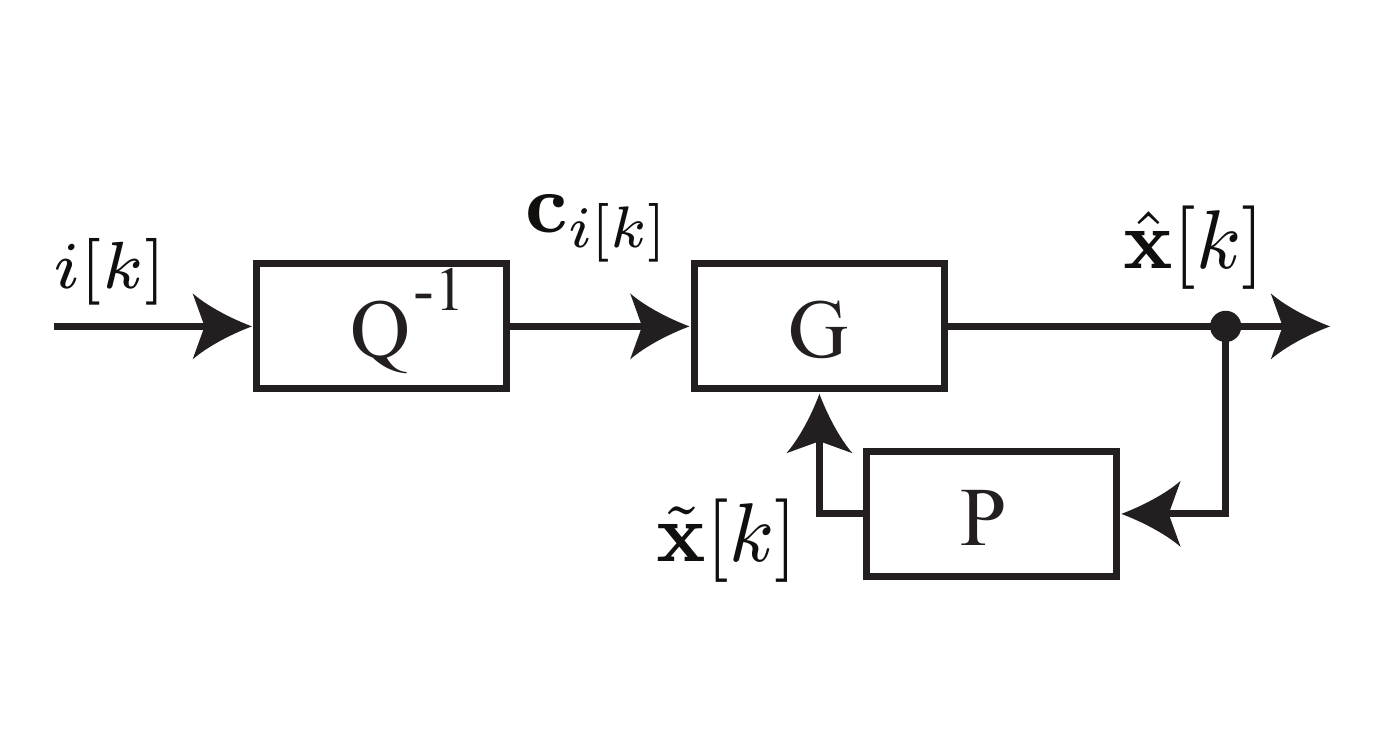}
\caption{Block diagram of predictive decoder on the Grassmann manifold.}
\label{fig:PVQ_Decoder}
\end{figure}

\figref{fig:PVQ_Decoder} illustrates the proposed GPC decoder; the pseudo code is shown in \algref{alg:GPCd}. The same error tangent codebook as the encoder is assumed to be available. The received indices are decoded in $Q^{-1}$ to recover $\bc_{i[k]}$. The predicted vector $\tilde{\bx}[k]$ is mapped to the estimated vector $\hat{\bx}[k]$ using the codeword as in \eqref{eq:estimate}. Similarly to the encoder, the prediction is performed using \eqref{eq:predict} to obtain $\tilde{\bx}[k+1]$ for the next time period. Note that for the first iteration of the decoder, the knowledge of $\tilde{\bx}[k]$, or equivalently $\hat{\bx}[k-1]$ and $\hat{\bx}[k-2]$, is needed. Synchronizing the initial vectors with the encoder is important because if $\tilde{\bx}[k]$ is different from the encoder, the received codeword no longer represents the correct error tangent vector. In Section \ref{sec:PC:initialization}, we provide an efficient strategy for initialization over finite rate communication channel. With appropriate initialization, symmetric operation at the encoder and decoder yields the same predicted vector $\tilde{\bx}[k]$ for each time $k$.

\subsection{Codebook Design}\label{sec:PC:codebook}
One of the strategy for PVQ codebook design is to employ an open loop approach followed by a closed loop approach to refine the codebook \cite{Gersho1991}. The open loop approach uses the prior vectors from a training data set to perform the prediction $\tilde{\bx}[k]=P(\bx[k-1],\bx[k])$ instead of predicting using the estimates, i.e. $P(\hat{\bx}[k-1],\hat{\bx}[k])$. The error tangent vector is computed using \eqref{eq:tan_k}. Then the Lloyd iterative algorithm is used to obtain the open-loop codebook. Using the codebook obtained using the open-loop codebook design, GPC is performed on the training data set to obtain a sequence of error tangent vectors. The Lloyd iteration is performed on the closed-loop error tangent vectors to obtain the final codebook. It is difficult to show the Lloyd iteration optimality of the open-loop and closed-loop approaches due to the feedback structure of the GPC but these approaches have been known to provide good results in the PVQ literature \cite{Gersho1991}. Thus, in this paper, we employ the open-loop and closed-loop approach to obtain the error tangent vector codebook. 

For storage of the codebook, we propose an efficient codebook representation by exploiting the product structure of the tangent space. We quantize separately the tangent vector magnitude and direction~\cite{Gersho1991}. Shape-gain vector quantization is widely used, for example, in speech and video coding~\cite{Sabin1984}. We use the shape-gain decomposition to provide efficient codebook storage and exploit it to analyze the rate-distortion of the proposed GPC that is otherwise very difficult. The tangent magnitude $\|\bee[k]\|$ is dependent on the distance between the predicted vector and the observed vector, which in turn is dependent on the rate of change of the input vectors. The unit norm error tangent vector depends on the location at which the tangent is computed and the directional statistics of the error. If $\cC$ is the obtained error tangent codebook of size $N_C$, the shape-gain decomposed codebooks are $\cC_d = \{\bc_{d,i}\}_{i=1}^{N_d}$ for the error tangent direction codebook and $\cC_m = \{c_{m,i}\}_{i=1}^{Nm}$ for the error tangent magnitude codebook. The desired codeword is reconstructed as $\bc_{i[k]} = c_{m,i[k]} \bc_{d,i[k]}$ at time $k$. With some heuristic design, it is possible to express, for example, a size $4$-bit codebook of vectors by a size $2$-bit codebook of scalars representing the magnitude and a size $2$-bit codebook of vectors representing the normalized tangent directions. Thus codebook storage reduction is possible at an expense of extra computation to reconstruct the codeword. 

\subsection{Initialization}\label{sec:PC:initialization}
Similar to the PVQ, the initial states of the GPC at both the encoder and the decoder needs to match to obtain the correct results. For example, in next generation wireless standards such as IEEE 802.16m, various feedback initialization intervals are defined \cite[Sec.16.3.6]{IEEE80216mD12}. Thus, an efficient mechanism for initialization is also important. Two approaches may be considered for initialization. One approach is to perform an initialization process so that the two estimated vectors $\hat{\bx}[k-1]$ and $\hat{\bx}[k-2]$ are communicated from the encoder to the decoder. Since the complete description of $\hat{\bx}[k-1]$ and $\hat{\bx}[k-2]$ must be communicated to the decoder, there is system dependent communication overhead. Another approach is to use the one-shot memoryless quantization technique to initialize the two vectors. This approach is attractive because it does not add any implementation overhead to systems already using one-shot feedback approach, e.g. 3GPP LTE. In particular, if the same codebook is used for the error tangent direction codebook and one-shot memoryless quantization codebook, there are no codebook memory overhead resulting in efficient implementation. A consequence of using memoryless quantization approach for initialization is that there may be an initial transient period in which the quantization error is larger than the steady state condition. As we show in Section \ref{sec:analysis}, this is because the memoryless quantization generally results in a larger quantization error.

\section{Performance Analysis of GPC}\label{sec:analysis}
In this section, we provide a quantization error analysis under a small angle approximation. We derive upper and lower distortion bounds, and then derive closed loop gain metric for the GPC algorithm. 

\subsection{Small Angle Approximation}
In this section we use the locally Euclidean property of the Grassmann manifold to derive an expression for the prediction error as a function of the tangent vector. If $\by \in \cG_{n,1}$ is obtained by changes to $\bx \in \cG_{n,1}$, we can approximate the chordal distance between $\bx$ and $\by$ as 
\begin{eqnarray}
  d(\bx, \by) & = & \sqrt{1 - | \bx^* \by |^2} \nonumber \\
  & = & |\sin(\theta)| \label{eq:chord_sine}\\
  & \approx & \|\bx - \by\| \label{eq:chord_2norm}
\end{eqnarray}
where \eqref{eq:chord_sine} follows from the subspace angle of vectors~\cite{Golub1996} and \eqref{eq:chord_2norm} follows from the small angle approximation. Thus, for a sufficiently small perturbation around $\bx$, the subspace distance between $\bx$ and $\by$ is approximated by the usual Euclidean distance. 

We may express the current observed vector at time $k$, $\bx[k]$, in terms of the predicted vector and the error tangent vector as
\begin{eqnarray}
  \bx[k] & = & G(\tilde{\bx}[k], \bee[k], 1) \nonumber \\
  & \approx & \tilde{\bx}[k] + \vec{\bee}[k] \|\bee[k]\| \nonumber \\
  & = & \tilde{\bx}[k] + \bee[k]
  \label{eq:obs_tan_vec}
\end{eqnarray}
using the small angle approximation. Furthermore, 
\begin{eqnarray}
  \bx^*[k] \bx[k] & \approx & ( \tilde{\bx}[k] + \bee[k] )^* ( \tilde{\bx}[k] + \bee[k] ) \nonumber \\
 & = & 1 + 2 \|\bee[k]\| \Re(\vec{\bee}[k]^* \tilde{\bx}[k]) + \|\bee[k]\|^2 \label{eq:ortho}\\
  & \approx & 1. \nonumber
\end{eqnarray}
The second term, $\vec{\bee}^*[k] \tilde{\bx}[k]$, in \eqref{eq:ortho} is zero because the unit norm tangent vector $\vec{\bee}[k]$ is orthogonal to $\tilde{\bx}[k]$. Similarly, if $\bc_{i[k]}$ is the selected error tangent codeword, the estimated signal can be expanded as
\begin{eqnarray}
  \hat{\bx}[k] & = & G(\tilde{\bx}[k], \bc_{i[k]}, 1) \nonumber \\
  & \approx & \tilde{\bx}[k] + \bc_{i[k]}.
  \label{eq:est_tan_vec}
\end{eqnarray}
Both \eqref{eq:obs_tan_vec} and \eqref{eq:est_tan_vec} reveal that for a small enough change, both vectors are expressed as an additive correction to the predicted vector. Thanks to the locally Euclidean property and using the usual $2$-norm for the local difference, the prediction error is
\begin{eqnarray}
  \|\bx[k] -  \hat{\bx}[k] \| 
  & \approx & \| \bee[k] - \bc_{i[k]} \|.
\end{eqnarray}
Therefore, the estimation error can be approximated as the normed difference between the actual tangent vector and the quantized tangent vector. Thus for small changes in the observed vector, the accuracy of tangent direction and tangent magnitude determines the accuracy of the estimate.

\subsection{Distortion Bounds}\label{sec:analysis:bounds}
The average distortion induced by a quantizer is a typical measure of performance. In what follows, we derive an upper and lower bound on the distortion for the proposed GPC algorithm. Recall that a metric ball $B_\delta(\bz)$ with radius $\delta$ centered at $\bz \in \cG_{n,1}$ on the Grassmann manifold is defined as 
\begin{equation}
  B_\delta(\bz) = \{ \by \in \cG_{n,1}: d(\by,\bz) \leq \delta\}
\end{equation}
such that $B_\delta(\bz) \subset \cG_{n,1}$. A closed form volume formula for $B_\delta(\bz)$ is given as~\cite{Dai2008}
\begin{equation}
  \text{Vol}(B_\delta(\bz)) = \delta^{2(n-1)}.
  \label{eq:vol_delta}
\end{equation}
Consider $B_\gamma(\bz) \subset \cG_{n,1}$ with $\delta \leq \gamma$ and volume of $B_\delta(\bz)$ given by \eqref{eq:vol_delta}. Let $(d\by)$ denote the differential form of the Haar measure on $\cG_{n,1}$. The distortion in the ball normalized by the volume of the ball was shown to be~\cite[Lemma 1]{Mondal2007}
\begin{equation}
  \int_{B_\gamma(\bz)} \frac{d^2(\by,\bz)(d\by)}{\text{Vol}(B_\gamma(\bz))} = \left( \frac{2(n-1)}{2n} \right) \gamma^2.
\end{equation}
For memoryless quantization, the volume together with a point density and covering assumption over the entire $\cG_{n,1}$ are used to characterize distortion. For the proposed GPC algorithm, the Voronoi region is determined by the tangent direction and tangent magnitude codebooks which makes the covering argument difficult. To overcome this difficulty, we assume that the tangent magnitude codebook provides concentric annular partitions of the sphere cap centered around the predicted vector and the tangent direction codebook partitioning each annulus into equiangle sectors. We obtain the bounds by considering the ball that is enclosed in the smallest annular sector and the ball that encloses the largest annular sector. Similarly, the distortion upper bound is given by the volume of the ball that covers the Voronoi cell.

Let $\gamma_d = \min_{\bc_{d,i},\bc_{d,k} \in \cC_d, i \neq k} d(\bc_{d,i},\bc_{d,k})$ denote the minimum chordal distance between the tangent direction codewords and $\gamma_m = \min_{c_{m,i},c_{m,k} \in \cC_m, i \neq k} | c_{m,i} - c_{m,k} |$ denote the minimum Euclidean distance between the tangent magnitude codewords. Similarly, let $\lambda_d = \max_{\bc_{d,i},\bc_{d,k} \in \cC_d, i \neq k} d(\bc_{d,i},\bc_{d,k})$ denote the maximum chordal distance between the tangent direction codewords and $\lambda_m = \max_{c_{m,i},c_{m,k} \in \cC_m, i \neq k} | c_{m,i} - c_{m,k} |$ denote the maximum Euclidean distance between the tangent magnitude codewords. Suppose that the tangent direction and magnitude codebooks maps uniformly to an equiangle sectors of concentric annulus centered at the predicted vector. Then the following lemma provides the bounds on the distortion for GPC algorithm.
\begin{lemma}[Distortion bounds]
If $\gamma_{\text{lower}} = \min \{\gamma_d, \gamma_m\}$ and $\lambda_{\text{upper}} = \max \{\lambda_d, \lambda_m\}$, lower and upper quantization distortion bounds are given by
\begin{eqnarray}
  D_{\text{lower}} & = & \left( \frac{2(n-1)}{2n} \right) \left( \frac{\gamma_{\text{lower}}}{2} \right)^2 \nonumber \\
  D_{\text{upper}} & = & \left( \frac{2(n-1)}{2n} \right) \left( \frac{\lambda_{\text{upper}}}{2} \right)^2.
\end{eqnarray}
\label{lemma:bound}
\end{lemma}
\begin{proof}
The lower bound is given by the volume of a metric ball that has ball radius which is smaller of the half minimum chordal distance of tangent direction codebook and half minimum distance of tangent magnitude codebook. The upper bound is similarly obtained by considering the volume of a metric ball which covers a Voronoi region. The bounds are exact since the metric ball volume formula is accurate~\cite[Lemma 1]{Mondal2007}.
\end{proof}
No claim is made on the tightness of the bound since an accurate description of the Voronoi region obtained by the proposed tangent codebook remains an open problem. In Section \ref{sec:simulation:bounds}, we provide numerical examples comparing the bounds obtained with actual distortion using fixed codebooks. 

Using the obtained lower bound, we may further quantify the reduction in distortion lower bound compared to memoryless quantization on the Grassmann manifold. For $\cG_{n,1}$, the lower bound on the fixed rate quantizer on the Grassmann manifold was shown to be
\begin{equation}
  D_{\cG_{n,1}}(N) = \left( \frac{2(n-1)}{2n} \right) N^{-\frac{1}{n-1}}
\end{equation}
where $N$ is the size of the codebook with rate $\log_2(N)$ bits~\cite{Dai2005,Mondal2007}. Suppose that $\gamma_{\text{lower}}$ is dominated by the tangent direction codebook such that $\gamma_{\text{lower}} = \gamma_d$ and that Grassmannian codebook is used for the tangent direction codebook. Then, the lower bound for the GPC algorithm can be expressed as
\begin{eqnarray}
  D_{\text{lower}} & = & \left( \frac{2(n-1)}{2n} \right) \left( \frac{\gamma_{\text{lower}}^2}{4} \right) \nonumber \\
  & = & \frac{1}{4} \left( \frac{2(n-1)}{2n} \right)^2 N_d^{-\frac{1}{n-1}} \nonumber \\
  & = & \frac{1}{4} \left( \frac{2(n-1)}{2n} \right) D_{\cG_{n,1}}(N_d)
  \label{eq:dist_reduction}
\end{eqnarray}
showing that the lower bound is smaller than $D_{\cG_{n,1}}(N_d)$ when $\gamma_d < \gamma_m$. 

\subsection{Performance Measures}
The closed loop prediction gain ratio is often used in vector quantization literature~\cite{Gersho1991} as a measure of how well the predictor performs with respect to the changes in the input. The closed loop prediction gain is usually written as the ratio of mean squared norm of the observed signal over mean squared norm of the prediction error. We define the mean squared error to be $\bbE[d^2(\tilde{\bx}[k], \bx[k])]$. For our GPC algorithm, we measure the closed loop prediction performance by
\begin{eqnarray}
   G_{\text{clp}} & = & \frac{\bbE[\|\bx[k]\|^2]}{\bbE[d^2(\tilde{\bx}[k], \bx[k])]} \nonumber \\
   & = & \frac{1}{\bbE[d^2(\tilde{\bx}[k], \bx[k])]}
  \label{eq:Grass_Gclp_general}
\end{eqnarray}
where $d^2(\tilde{\bx}[k], \bx[k])$ denotes the squared chordal prediction error. In fact, \eqref{eq:Grass_Gclp_general} can be further expressed as a function of the tangent vector assuming that the small angle approximation holds. Using \eqref{eq:chord_2norm} and \eqref{eq:obs_tan_vec}, the distance function in the denominator can be approximated as $d(\tilde{\bx}[k],\bx[k]) \approx \| \bee[k] \|$. Therefore, the closed loop prediction gain for GPC algorithm becomes
 \begin{eqnarray}
  G_{\text{clp}} & \approx & \frac{1}{\bbE[ \|\bee[k]\|^2 ]}
  \label{eq:Grass_Gclp}
\end{eqnarray}
which shows the dependence of the closed loop prediction gain performance on the tangent magnitude. The tangent magnitude is in turn dependent on the changes in the observed process. A closed form relationship between the observed process and the tangent magnitude is in general difficult to obtain. In Section \ref{sec:simulation:gains}, we show some empirical results of the closed loop prediction gain performance for the proposed GPC algorithm. 

\section{Simulation Results}\label{sec:simulation}
In this section, we provide numerical results to illustrate the performance of the proposed GPC algorithm. 

\subsection{Distortion Bounds}\label{sec:simulation:bounds}
\begin{figure}[!htbp]
\centering
\includegraphics[width=0.9\columnwidth]{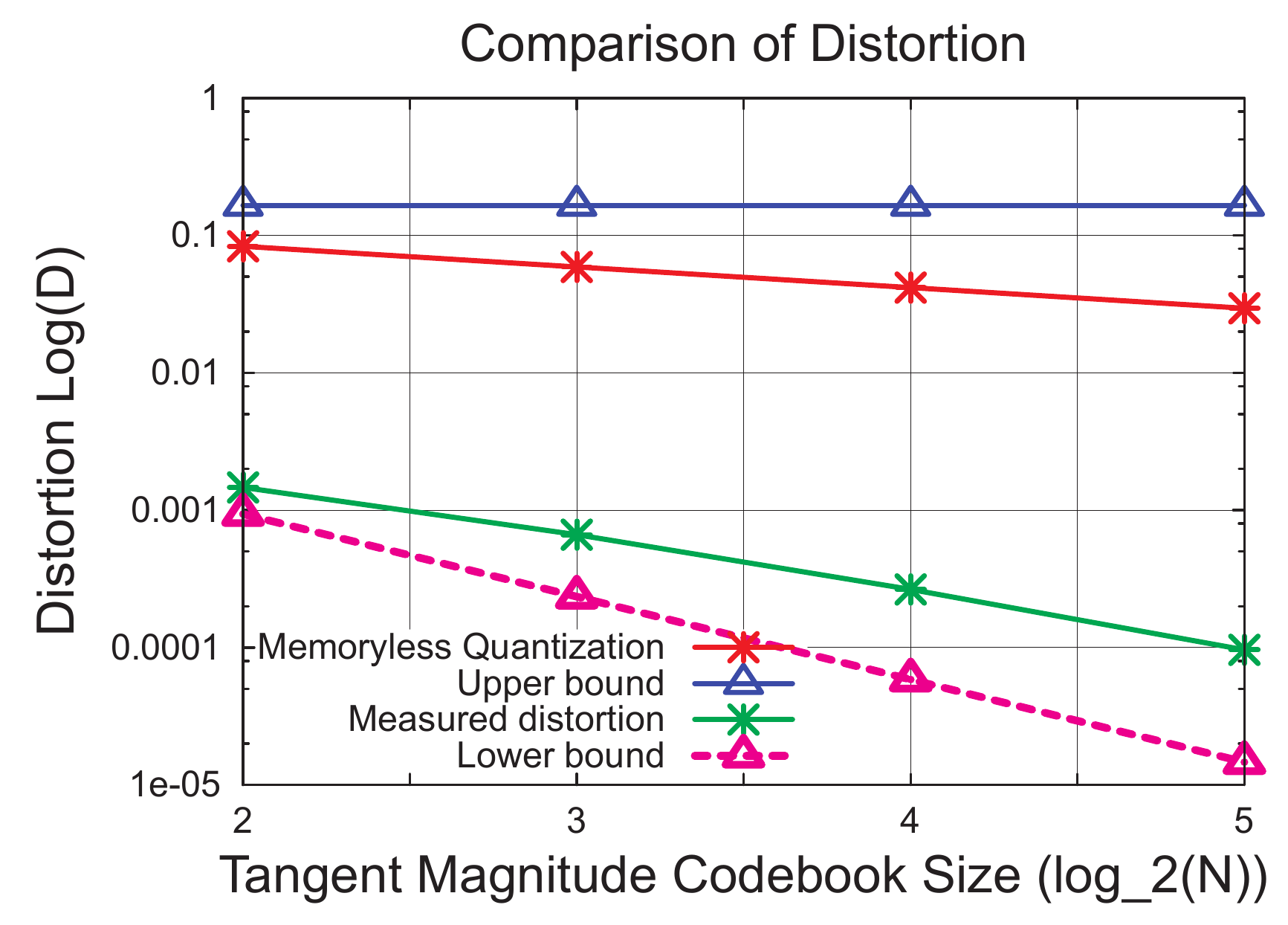}
\caption{Comparison of operational distortion against the lower and upper bound for various tangent magnitude codebook size. The lower distortion bound for the memoryless quantizer using Grassmannian codebook is also shown.}
\label{fig:distbound}
\end{figure}

We present a numerical example illustrating the operational distortion and compare it with the upper and lower bounds given in Lemma \ref{lemma:bound}. Correlated $3 \times 1$ vectors were generated according to a second order autoregressive model with memory coefficients $\alpha_1 = 0.9$ and $\alpha_2 = 0.75$ with additive noise distributed according to zero mean complex Gaussian with variance $(0.01)^2$, {\it i.e.,}  $\bh[k] = \alpha_1 \bh[k-1] + \alpha_2 \bh[k-2] + \sqrt{1-\alpha_1^2 - \alpha_2^2} \bz[k]$. The normalized vectors were considered to be the samples on $\cG_{3,1}$ to which the proposed GPC algorithm was applied. For this experiment, an $N_d = 2^4$ tangent direction codebook was used and the tangent magnitude codebook size was varied from $N_m = 2^2$ to $2^5$. \figref{fig:distbound} shows the operational distortion with upper and lower distortion bounds obtained in Lemma \ref{lemma:bound} as a function of the tangent magnitude codebook size. The lower bound captures the distortion trend over the range of codebook sizes while the upper bound seems too loose. We also illustrate the lower bound of a memoryless quantization using a Grassmannian codebook with codebook sizes of $6$, $7$, $8$, and $9$ bits so that the total number of bits used for the codebook matches that of the proposed GPC algorithm. We see that the proposed GPC algorithm provides significant improvement in distortion over the memoryless quantization technique. Unfortunately, the upper bound from Lemma \ref{lemma:bound} is dominated by the resolution of the $4$-bit tangent direction codebook which has higher distortion than the memoryless quantization with adjusted number of codebook size. Nevertheless, the result shows that a significant reduction in distortion is achieved by the proposed GPC algorithm and the achievable distortion can be controlled by the tangent magnitude codebook which is a simple scalar codebook. 

\subsection{Closed Loop Prediction Gain and Prediction Error}\label{sec:simulation:gains}
\begin{figure}[!htbp]
\centering
\includegraphics[width=0.9\columnwidth]{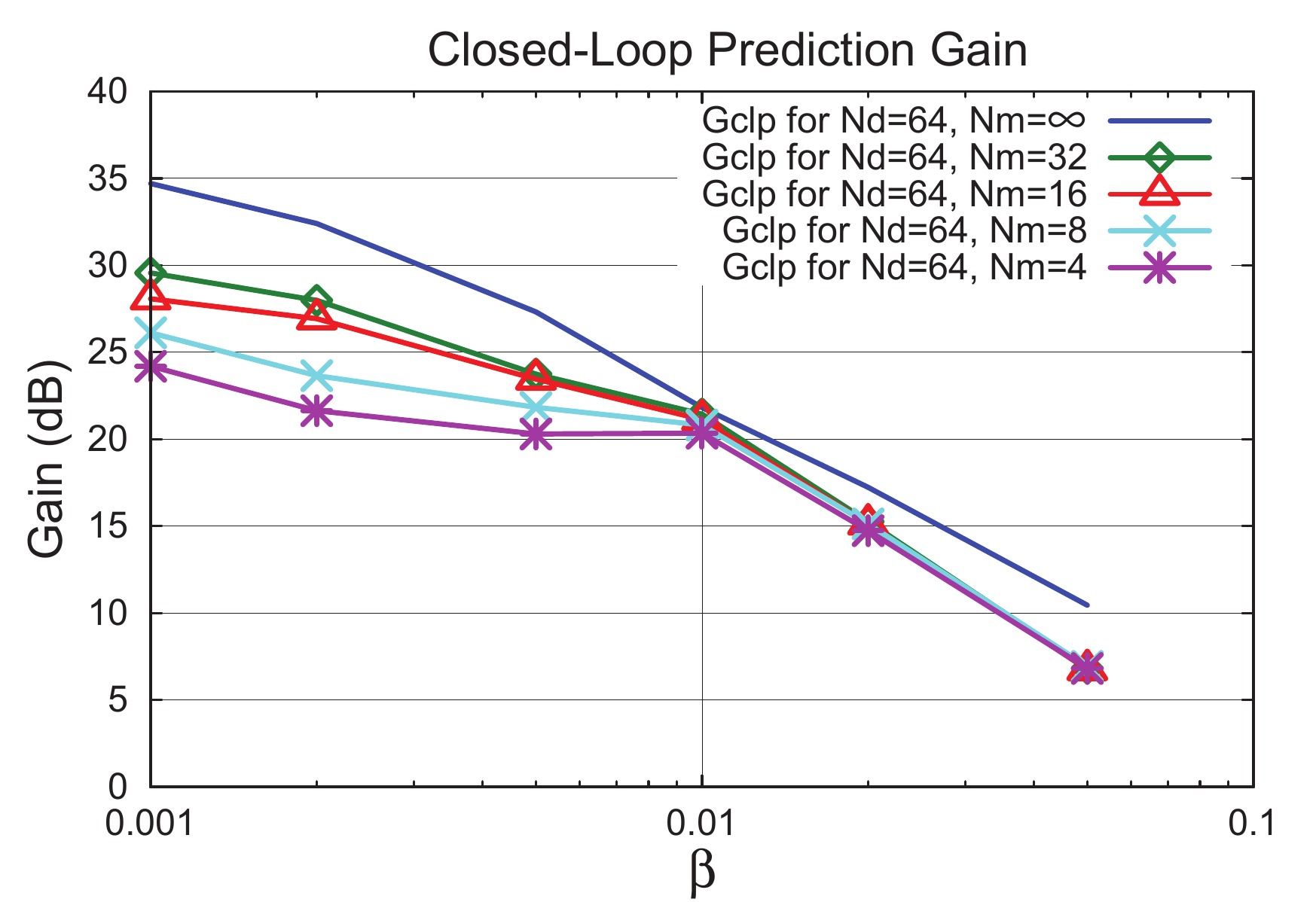}
\caption{Closed loop prediction gain, $G_{\text{clp}}$, for $\cG_{4,1}$ data with fixed tangent codebook ($N_d=64$) and different tangent magnitude codebooks ($N_m = 2^2,2^3,2^4,2^5$) over various correlation parameter $\beta$.}
\label{fig:gains3}
\end{figure}

To illustrate the dependence on the tangent direction and tangent magnitude codebooks, \figref{fig:gains3} shows the closed loop prediction gains for various error tangent magnitude codebook sizes and fixed tangent direction codebook of size $N_d=64$. For these numerical examples, a correlated $4 \times 1$ vector sequence was generated according to a first order autoregressive model (or Gauss-Markov model~\cite{Etkin2006}) with correlation coefficient $\alpha = J_0 (2 \pi \beta)$ where $J_0$ is Bessel function of zeroth order and $\beta$ is the normalized Doppler frequency. The sequence of channel coefficients are generated according to
\begin{equation}
  \bh[k] = \alpha \bh[k-1] + \sqrt{1-\alpha^2} \bz[k]
  \label{AR1}
\end{equation}
where $k$ is the time index and $\bz[k]$ is a $\Nt \times 1$ vector with each entry drawn from an i.i.d. zero mean complex white Gaussian process. The normalized vectors $\bx[k] = \bh[k] / \|\bh[k]\|$ are the correlated sequence on the Grassmann manifold. For the tangent direction codebook, an $N_d=2^6$ Grassmannian codebook~\cite{LoveCodebook} was used and the tangent magnitude codebooks were based on a uniform quantization between $0$ and $1$ using $2$, $3$, $4$, and $5$ bits. For an upper bound, the closed loop prediction gain without quantizing the tangent magnitude is also shown. The result illustrates the dependence of closed loop prediction gain on tangent magnitude codebook size as a function of correlation parameter $\beta$. For highly correlated data, the tangent magnitude codebook resolution has higher impact on the closed loop prediction gain. This is because the smallest tangent magnitude quantization level may be larger than the prediction error leading to an over estimation. If the tangent magnitude codebook is adjusted based on the correlation, e.g., quantize in the range of $[0,0.1]$ instead of $[0,1]$, this gap may be closed. 

\begin{figure}[!htbp]
\centering
\includegraphics[width=0.9\columnwidth]{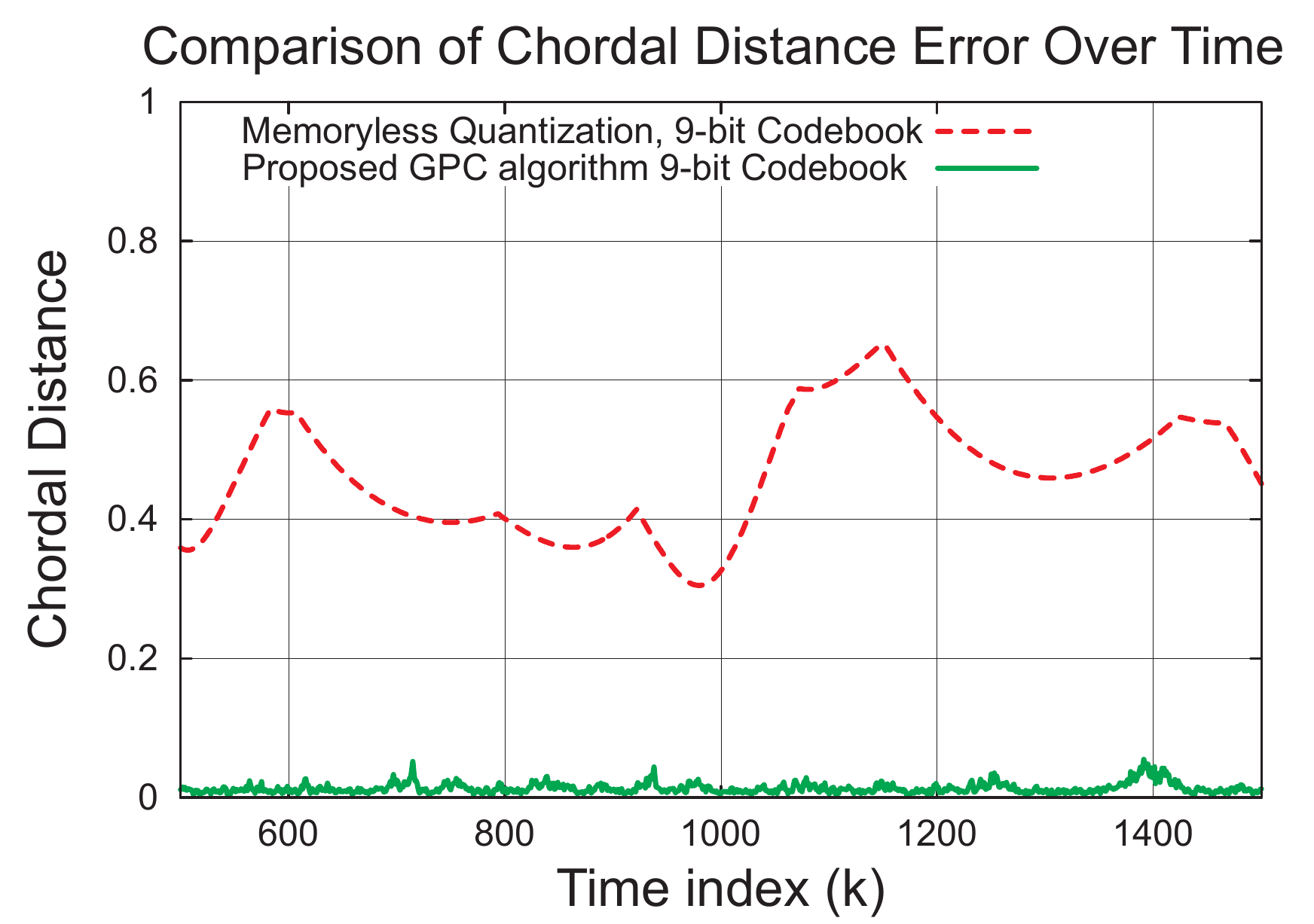}
\caption{Chordal distance comparison over time between memoryless quantizer using $9$-bit Grassmannian codebook and the proposed GPC algorithm using $9$-bit codebook.}
\label{fig:error}
\end{figure}

Another useful performance measure is the chordal distance error between the estimated vector $\hat{\bx}[k]$ and the observed vector $\bx[k]$. The chordal distance error $d(\hat{\bx}[k], \bx[k])$ shows how close the estimated vector is to the observed vector using the proposed GPC algorithm. In MIMO communication application considered in \ref{sec:MUMIMO}, the chordal distance error has a direct impact on the respective communication theoretic performance measures. In \figref{fig:error}, we show the chordal distance between $\hat{\bx}[k]$ and $\bx[k]$ and the chordal distance between the quantized vector and the observed vector for memoryless quantization using Grassmannian codebook with $N=2^9$. \figref{fig:error} illustrates the substantial improvement in the quantization accuracy compared with memoryless technique. 

\begin{figure}[!htbp]
\centering
\includegraphics[width=0.9\columnwidth]{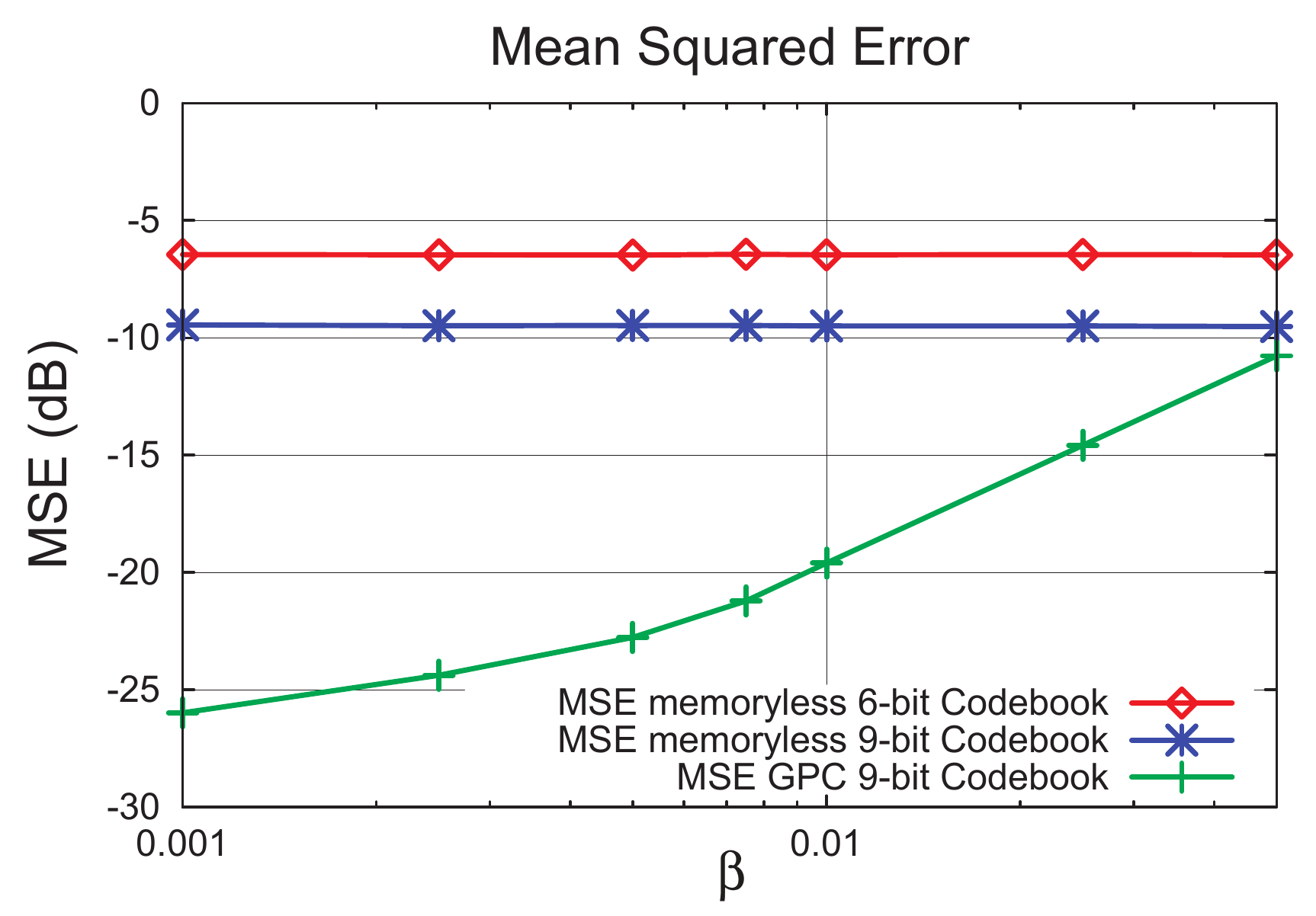}
\caption{Mean squared error comparison between memoryless quantization using $6$-bit and $9$-bit Grassmannian codebook and the proposed GPC algorithm using $9$-bit codebook.}
\label{fig:MSE}
\end{figure}

To further illustrate the quantizer accuracy, we show the operational mean squared chordal distance error (MSE) as a function of $\beta$ for the proposed GPC algorithm and memoryless quantizer using Grassmannian codebook in \figref{fig:MSE}. The memoryless quantizer provides approximately $-7$ dB of MSE whereas the proposed GPC algorithm provides as little as $-26$ dB of MSE which shows that significant accuracy can be obtained over memoryless quantization techniques. As the correlation decreases, the MSE approaches that of the memoryless quantization MSE. 

\subsection{Application to Zero Forcing Multiuser MIMO System}\label{sec:MUMIMO}
In this section, we illustrate the application of proposed GPC algorithm to limited feedback multiuser MIMO system using zero forcing precoding~\cite{Jindal2006}. We assume that the transmitter has $\Nt = 4$ transmit antennas and each user is equipped with single receive antenna. We assume that the encoder and decoder are initialized and that each user has a perfect channel estimate. Then, each user performs the prediction as described in Section \ref{sec:PC} and feedback the indices of quantized tangent direction and tangent magnitude codewords. The transmitter uses the received indices and performs the prediction as depicted in \figref{fig:PVQ_Decoder}. Then, the predicted channel vectors are used to form the composite channel matrix to compute the zero forcing precoder. The channel to each user is assumed to be temporally correlated with correlation according to $J_0 (2 \pi f_D T_s)$ \cite{Simon2007}. Each user's channel is independently generated assuming same temporal correlation. 

To compare the random codebook approach and the proposed GPC algorithm, we compare the achievable sum rate for three scenarios. First, the achievable sum rate assuming perfect CSI at the transmitter is obtained. For the perfect CSI case, i.i.d. channel is assumed. The perfect CSI case provides the baseline for what can be achieved. The second scenario is the random vector codebook approach also assuming i.i.d. channel \cite{Jindal2006}. Finally, the proposed GPC algorithm using 9-bit codebook for $f_DT_s= 0.001, 0.01, 0.02,$ and $0.04$ that corresponds to Doppler frequencies of 0.2Hz, 2Hz, 4Hz, and 8 Hz at 5ms update intervals that is found in LTE-Advanced and IEEE 802.16m. 

\begin{figure}[!htbp]
\centering
\includegraphics[width=0.9\columnwidth]{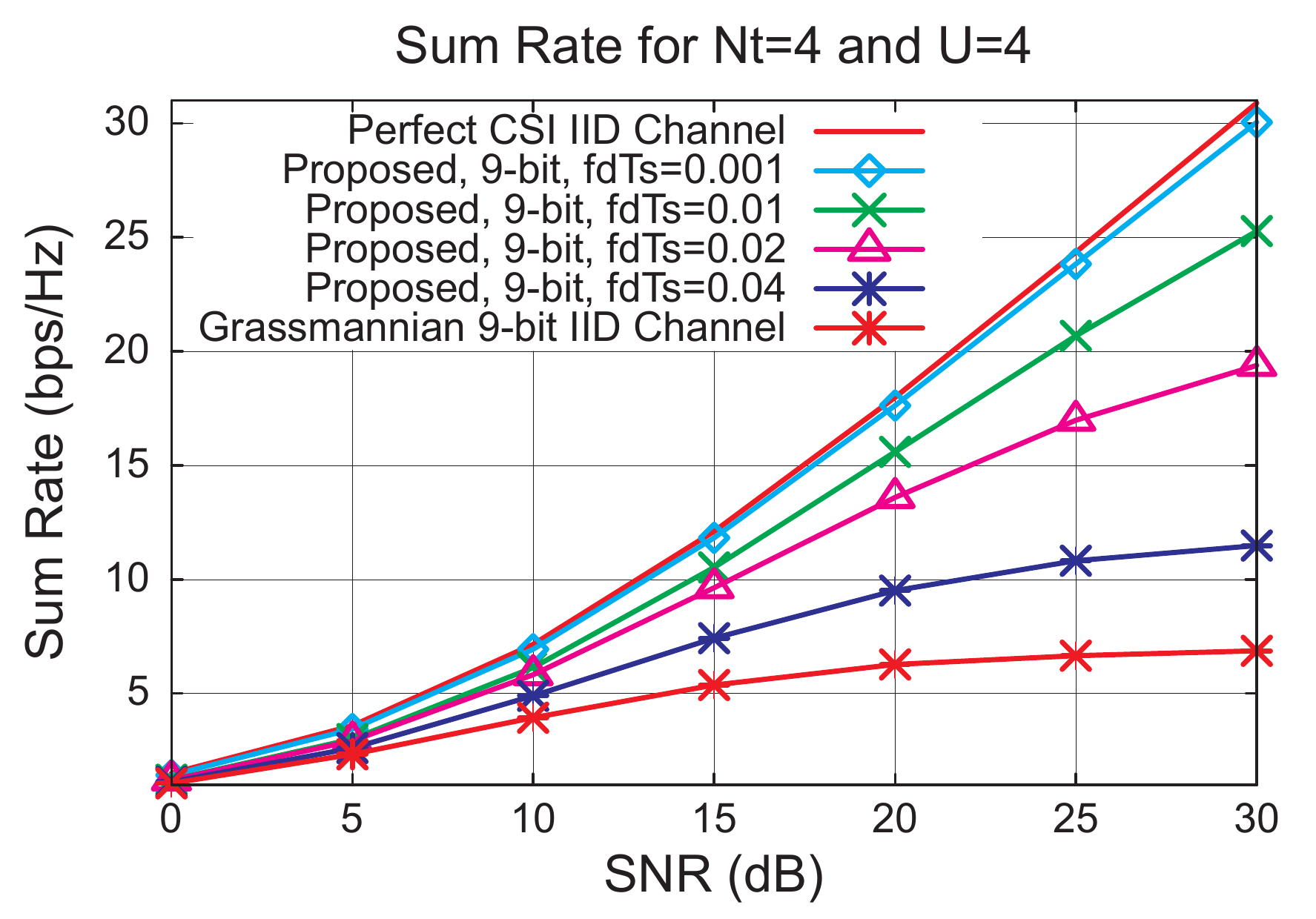}
\caption{Sum rate for $\Nt=U=4$ i.i.d. channel with perfect CSI, i.i.d. channel with $9$-bit Grassmannian codebook, and the proposed GPC algorithm with $9$-bit codebook for various normalized Doppler frequencies.}
\label{fig:MU4Cap}
\end{figure}

\figref{fig:MU4Cap} illustrates the achievable sum rate for cases being considered. Contrary to the random codebook strategy, the proposed GPC algorithm provides significant sum rate gain. In fact, for $f_D T_s = 0.001$, the system starts to become interference limited above SNR of 20dB illustrating the superior CSI accuracy when the channel is highly correlated. Furthermore, each user is equipped with the same codebooks which eliminates the need to store multiple codebooks at the transmitter, thus reducing the overhead for practical applications. 

\begin{figure}[!htbp]
\centering
\includegraphics[width=0.9\columnwidth]{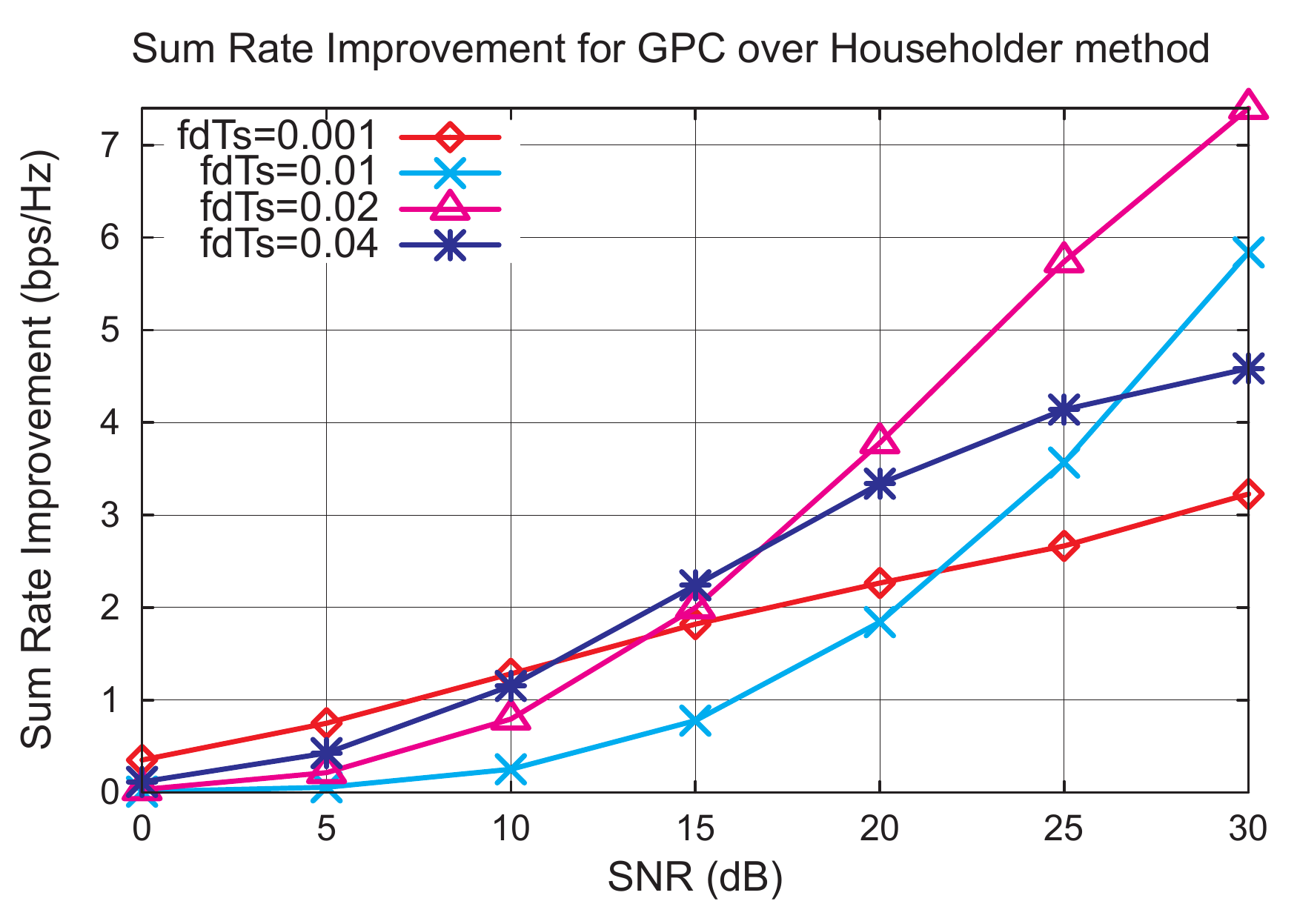}
\caption{Sum rate improvement obtained by the proposed GPC algorithm over Householder technique for $\Nt=U=4$ using $9$-bit codebook for various normalized Doppler frequencies.}
\label{fig:MU4Compare}
\end{figure}

\figref{fig:MU4Compare} illustrates the sum rate improvement of the proposed technique over the Householder technique in \cite{Liu2006a} over a range of SNR for channels with various normalized Doppler frequencies. Both methods used $9$-bit feedback per channel use. The plot shows that the proposed GPC algorithm outperforms the Householder technique especially at high SNR illustrating higher CSI resolution obtained by the GPC algorithm. 

\section{Conclusion}\label{sec:conclusion}
In this paper, we proposed a new predictive coding algorithm on the Grassmann manifold for limited feedback in multiple antenna wireless systems. Building on the classical predictive vector quantization on linear vector space and the geometric properties of the Grassmann manifold, we derived a predictive coding framework for $\cG_{n,1}$. Distortion bounds were obtained showing possible distortion improvement over memoryless quantization technique. In simulations we showed that the proposed GPC algorithm provides significant sum rate improvement for multiuser MIMO system using practical codebook size. Future work should consider the optimization of the tangent magnitude codebook and extensions to a higher dimensional Grassmann manifold, i.e., $\cG_{n,p}$ for $p>1$.

\appendices

\section{Proof of Lemma \ref{lemma:tangent}}\label{app:lemma:tangent}
\begin{proof}
It was shown in \cite{Inoue2009} that the tangent vector between $\bx_1$ and $\bx_2$ in $\cG_{n,1}$ can be written as
\begin{equation}
  \bee = \tan^{-1} \left( \left\| \frac{\bx_2}{\rho} - \bx_1 \right\| \right) \frac{\bx_2 / \rho - \bx_1}{\|\bx_2 / \rho - \bx_1\|}.
\end{equation}
The normed term can be simplified as
\begin{eqnarray}
  \left\| \frac{\bx_2}{\rho} - \bx_1 \right\|^2 & = & \left(\frac{\bx_2}{\rho} - \bx_1 \right)^* \left(\frac{\bx_2}{\rho} - \bx_1\right) \nonumber \\
  & = & \frac{1}{|\rho|^2} - 1.
\end{eqnarray}
Therefore, 
\begin{eqnarray}
  \left\| \frac{\bx_2}{\rho} - \bx_1 \right\| & = & \sqrt{ \frac{1}{|\rho|^2} - 1}  = \frac{d}{|\rho|} \nonumber 
\end{eqnarray}
where $d = \sqrt{1-|\rho|^2}$ is the chordal distance between $\bx_1$ and $\bx_2$. Clearly, $\|\bee\| = \tan^{-1} (d/\|\rho|) \geq 0$ and $\vec{\bee} = (\bx_2 / \rho - \bx_1)/(d/|\rho|)$ such that $\bee = \|\bee\| \vec{\bee}$. 

Using the exponential form of trigonometric identities $\tan^{-1}(x) = (j/2) \ln \{ (1 - jx)/(1+jx) \}$ and $\cos^{-1}(x) = -j \ln (x + \sqrt{x^2 - 1})$, we have
\begin{eqnarray}
  \tan^{-1} \left( \frac{d}{|\rho|} \right) & = & \frac{j}{2} \ln \left( \frac{1 - j \left( \frac{d}{|\rho|^2} \right)}{1 + j \left( \frac{d}{|\rho|^2} \right)}\right) \nonumber \\
  & = & -j \ln (|\rho| + \sqrt{|\rho|^2 - 1}) \nonumber \\
  & = & \cos^{-1} |\rho|.
\end{eqnarray}
Since $|\rho|$ is the cosine of the subspace angle between $\bx_1$ and $\bx_2$, this shows that the norm of the tangent vector is equal to the arc length, i.e., $| \theta |$ with subspace angle $\theta$ \cite[p. 603]{Golub1996}. 
\end{proof}

\section{Proof of Lemma \ref{lemma:geodesic}}\label{app:lemma:geodesic}
\begin{proof}
  For the general case where $\bX_1,\ \bX_2 \in \cG_{n,p}$ with $n > p > 0$, the geodesic between $\bX_1$ and $\bX_2$ was shown to be~\cite{Edelman1998}
  \begin{eqnarray}
    \bX(t) & = & \bX_1 \bV \cos (\Sigma t) \bV^* + \bU \sin(\Sigma t) \bV^* \nonumber
  \end{eqnarray}
  where $\bU\Sigma\bV^*$ is the compact singular value decomposition of the tangent emanating from $\bX_1$ to $\bX_2$. For the case $\bx_1,\ \bx_2 \in \cG_{n,1}$, let $\bee$ be the tangent vector emanating from $\bx_1$ to $\bx_2$. Then, we may assume $\bV = 1$ without loss of generality and identify $\bU$ with $\vec{\bee}$ and $\Sigma$ with $\|\bee\|$ to obtain 
  \begin{equation} 
    G(\bx_1, \bee, t) = \bx_1 \cos(\|\bee\| t) + \vec{\bee} \sin(\|\bee\| t).
  \end{equation} 
  It is clear  that $G(\bx_1,\bee,0) = \bx_1$. At $t=1$, we have
  \begin{eqnarray}
    G(\bx_1, \bee, 1) & = & \frac{\bx_1}{\sqrt{1+ d^2/|\rho|^2}} \nonumber \\
    & & + \frac{\bx_2/\rho - \bx_1}{d/|\rho|} \frac{d/|\rho|}{\sqrt{1+ d^2/|\rho|^2}} \label{pf:geod:1} \\
    & = & \frac{\bx_2}{\rho \sqrt{1+ d^2/|\rho|^2}} \label{pf:geod:2} \\
    & = & \bx_2 \nonumber 
  \end{eqnarray}
  where we have used the identities
  \begin{eqnarray}
    \sin(x) & = & \frac{x}{\sqrt{1+x^2}} \nonumber \\
    \cos(x) & = & \frac{1}{\sqrt{1+x^2}} \label{pf:trig:sincos}
  \end{eqnarray}
  in \eqref{pf:geod:1} and the fact that $\rho\sqrt{1+ d^2/|\rho|^2} = 1$ in \eqref{pf:geod:2}. 

  To verify that $G(\bx_1,\bee, t)$ for $t \in [0,1]$ is a valid point on the Grassmann manifold, taking the inner product of $G(\bx_1,\bee,t)$ with itself yields $1$ for $t \in [0,1]$ by using the fact that $\bx_1 \perp \bee$.
\end{proof}

\section{Proof of Lemma \ref{lemma:pt}}\label{app:lemma:pt}
\begin{proof}
  For the general case where $\bX_1,\ \bX_2 \in \cG_{n,p}$, $n > p > 0$, the parallel transport of tangent $\bE$ emanating from $\bX_1$ along the geodesic direction $\Delta$ with compact singular value decomposition, $\bU \Sigma \bV^*$, was shown to be~\cite{Edelman1998}
  \begin{equation}
    \hat{\bE} = \left[ -\bX_1 \bV \sin(\Sigma t) \bU^* + \bU \cos(\Sigma t) \bU^* + (\bI - \bU \bU^*) \right] \bE.
  \end{equation}
We need to show the parallel transport of the tangent vector $\bee$ emanating from $\bx_1$ to $\bx_2$ in the geodesic direction $\bee$ for the case $\bx_1,\ \bx_2 \in \cG_{n,1}$. Without loss of generality, we assume that the singular value decomposition of $\bee$ is given with $\vec{\bee}$ as the left singular vector, $\|\bee\|$ as the singular value, and $1$ for the right singular vector. Then
  \begin{eqnarray}
    \vec{\bee}(t) & = & \left[ -\bx_1 \vec{\bee}^* \sin(\|\bee\| t) + \vec{\bee} \vec{\bee}^* \cos(\|\bee\| t) + (\bI - \vec{\bee} \vec{\bee}^*) \right] \bee \nonumber \\
    & = & -\bx_1 \|\bee\| \sin(\|\bee\| t) + \bee \cos(\|\bee\| t). \label{pf:pt:1}
  \end{eqnarray}
  Since $G(\bx_1, \bee, 1) = \bx_2$, the parallel transported tangent vector emanating from $\bx_2$ is found by evaluating \eqref{pf:pt:1} for $t=1$. Using \eqref{eq:tangent} and \eqref{pf:trig:sincos}, we have
  \begin{eqnarray}
    \hat{\bee} & = & -\bx_1 \|\bee\| \sin(\|\bee\|) + \bee \cos(\|\bee\|) \nonumber \\
    & = & \frac{-\bx_1 \tan^{-1} (d/|\rho|) (d/|\rho|)}{\sqrt{1 + d^2 / |\rho|^2}} \nonumber \\
    &  & + \frac{\tan^{-1}(d/|\rho|) (\bx_2 / \rho - \bx_1)}{d/|\rho|} \frac{1}{\sqrt{1+d^2 / |\rho|^2}} \nonumber \\
    & = & \frac{\tan^{-1}(d/|\rho|)}{(d/|\rho|)\sqrt{1 + d^2 / |\rho|^2}} \left( \frac{\bx_2}{\rho} - \bx_1 \left( 1+ \frac{d^2}{|\rho|^2} \right) \right) \nonumber \\
    & = & \tan^{-1} \left( \frac{d}{|\rho|} \right) \frac{\bx_2 \rho^* - \bx_1}{d}
  \end{eqnarray}
  which is the desired result.
\end{proof}

\section{Derivation of Prediction Function in Definition \ref{def:prediction}}\label{app:def:prediction}

  Recall that the parallel transported tangent vector $\hat{\bee}$ emanating from $\bx_2$ is given in \eqref{eq:new_pt}. Computing the geodesic with from $\bx_2$ along $\hat{\bee}$ at $t=1$ gives
  \begin{eqnarray}
    \hat{\bx} & = & G(\bx_2, \hat{\bee}, 1) \nonumber \\
    & = & \frac{\bx_2}{\sqrt{1 + d^2 / |\rho|^2}} + \frac{\bx_2 \rho^* - \bx_1}{\sqrt{1 + d^2 / |\rho|^2}} \nonumber \\
    & = & |\rho| \bx_2 + \rho^* \bx_2 - \bx_1.
  \end{eqnarray}
  To see that $\hat{\bx} \in \cG_{n,1}$, we have
  \begin{eqnarray}
    \hat{\bx}^* \hat{\bx} & = & (|\rho| \bx_2 + \rho^* \bx_2 - \bx_1)^* (|\rho| \bx_2 + \rho^* \bx_2 - \bx_1) \nonumber \\
    & = & 1
  \end{eqnarray}
  where we have used the fact that $\rho = \bx_1^* \bx_2$. To see that the prediction is distance preserving, the inner product of $\bx_2$ and $\hat{\bx}$ gives
  \begin{eqnarray}
    \bx_2^* \hat{\bx} & = & \bx_2^* \bx_2 |\rho| + \bx_2^* \bx_2 \rho^* - \bx_2^* \bx_1 \nonumber \\
    & = & |\rho|. 
  \end{eqnarray}
  Therefore, 
  \begin{equation}
    d(\bx_2, \hat{\bx}) = \sqrt{1 - |\rho|^2} = d(\bx_1, \bx_2).
  \end{equation}

\ifCLASSOPTIONcaptionsoff
  \newpage
\fi

\input{Predict2011TwoColumn.bbl}

\begin{algorithm}
\caption{GPC encoder algorithm}
\label{alg:GPCe}
\begin{algorithmic}[1]
\REQUIRE $\bx[k]$
\STATE Initialize $\tilde{\bx}[1]$ and $\hat{\bx}[0]$
\FORALL{k=1,2,\dots}
  \STATE $\bee[k] = L(\tilde{\bx}[k], \bx[k])$
  \STATE $i[k] = Q(\bee[k])$
  \STATE $\hat{\bx}[k] = G(\tilde{\bx}[k], \bc_{i[k]}, 1)$
  \STATE $\tilde{\bx}[k+1] = P(\hat{\bx}[k-1],\hat{\bx}[k])$
\ENDFOR
\ENSURE $i[k]$
\end{algorithmic}
\end{algorithm}

\begin{algorithm}
\caption{GPC decoder algorithm}
\label{alg:GPCd}
\begin{algorithmic}[1]
\REQUIRE $i[k]$
\STATE Initialize $\tilde{\bx}[1]$ and $\hat{\bx}[0]$
\FORALL{k=1,2,\dots}
  \STATE $\bc_{i[k]} = Q^{-1}(i[k])$
  \STATE $\hat{\bx}[k] = G(\tilde{\bx}[k], \bc_{i[k]}, 1)$
  \STATE $\tilde{\bx}[k+1] = P(\hat{\bx}[k-1],\hat{\bx}[k])$ 
\ENDFOR
\ENSURE $\hat{\bx}[k]$
\end{algorithmic}
\end{algorithm}


\begin{figure}[!htbp]
\centering
\includegraphics[width=0.9\columnwidth]{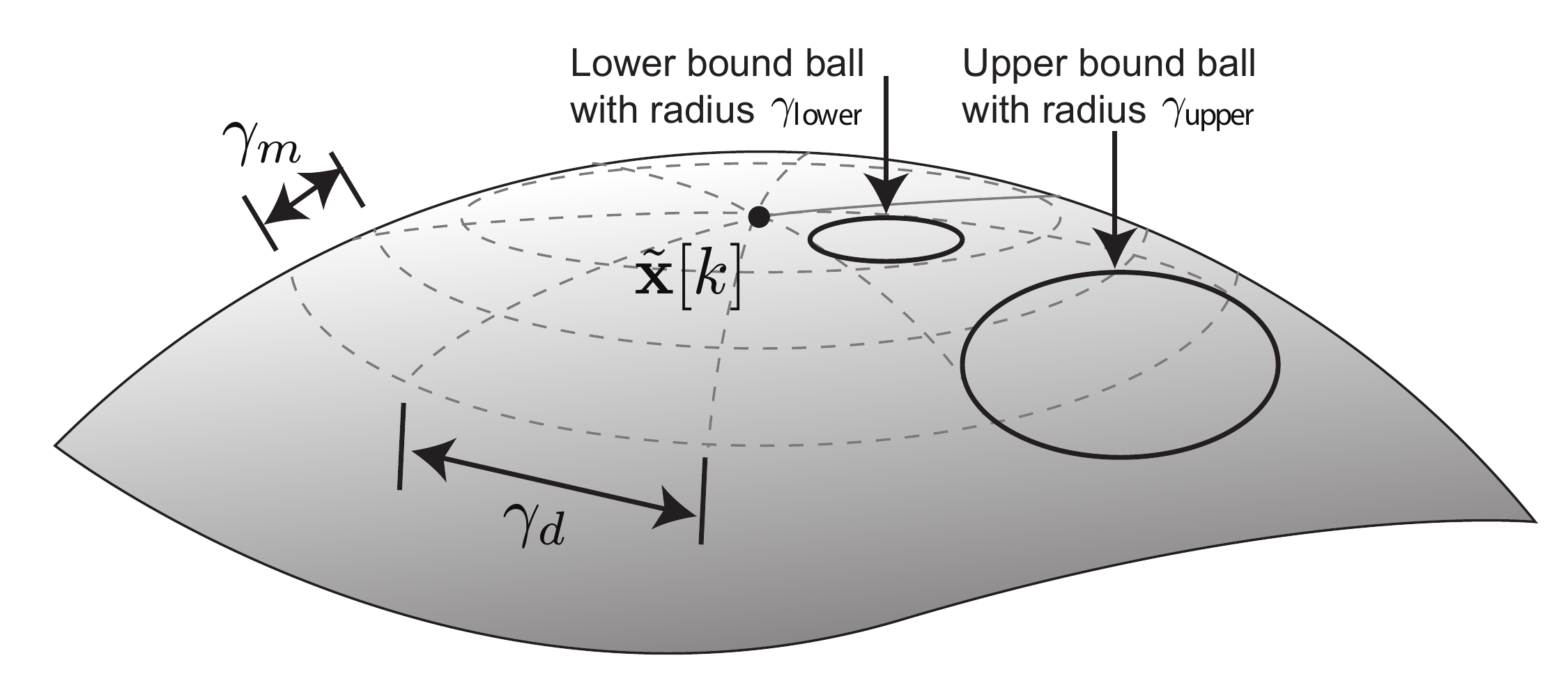}
\caption{Illustration of quantization region around the predicted vector and lower and upper distortion bound balls.}
\label{fig:Ball}
\end{figure}

\end{document}

%% file: header.tex
\usepackage{cite}

\ifCLASSINFOpdf
  \usepackage[pdftex]{graphicx}
  \graphicspath{{figures/}}
\else
\fi

\usepackage[cmex10]{amsmath}
\usepackage{array}

\usepackage{mdwmath}
\usepackage{mdwtab}

\usepackage[tight,footnotesize]{subfigure}

\usepackage[font=footnotesize]{subfig}
\usepackage{fixltx2e}

\ifCLASSOPTIONcaptionsoff
  \usepackage[nomarkers]{endfloat}
 \let\MYoriglatexcaption\caption
 \renewcommand{\caption}[2][\relax]{\MYoriglatexcaption[#2]{#2}}
\fi
\usepackage{url}

\usepackage{dsfont}
\usepackage{yfonts}

%% file: input.tex
\usepackage{amsfonts}
\usepackage{times}
\usepackage{latexsym}
\usepackage{amssymb}
\usepackage{amsmath}
\usepackage{cite}
\usepackage{algorithm}
\usepackage{algorithmic}

\newtheorem{theorem}{Theorem}

\newtheorem{definition}[theorem]{Definition}

\newtheorem{lemma}[theorem]{Lemma}

\newenvironment{proof}{\begin{quote}}{ \end{quote} \hfill $\Box$}
  
\newcommand{\figref}[1]{{Fig.}~\ref{#1}}

\newcommand{\algref}[1]{{Algorithm}~\ref{#1}}

\DeclareMathOperator*{\argmin}{arg\,min}

\newcommand{\bI}{\mathbf{I}}

\newcommand{\cN}{\mathcal{N}}

\def\Nt{{N_t}}

\def\bydef{:=}

\def\bb0{{\mathbb{0}}}

\def\bb{{\mathbf{b}}}
\def\bc{{\mathbf{c}}}

\def\bee{{\mathbf{e}}}

\def\bg{{\mathbf{g}}}
\def\bh{{\mathbf{h}}}

\def\bs{{\mathbf{s}}}

\def\bv{{\mathbf{v}}}

\def\bx{{\mathbf{x}}}
\def\by{{\mathbf{y}}}
\def\bz{{\mathbf{z}}}
\def\b0{{\mathbf{0}}}

\def\bA{{\mathbf{A}}}

\def\bE{{\mathbf{E}}}

\def\bH{{\mathbf{H}}}
\def\bI{{\mathbf{I}}}

\def\bU{{\mathbf{U}}}
\def\bV{{\mathbf{V}}}

\def\bX{{\mathbf{X}}}

\def\bbC{{\mathbb{C}}}

\def\bbE{{\mathbb{E}}}

\def\bbN{{\mathbb{N}}}

\def\cC{\mathcal{C}}

\def\cG{\mathcal{G}}

\def\cN{\mathcal{N}}

\def\cR{\mathcal{R}}

\def\cU{\mathcal{U}}

\def\sf0{{\mathsf{0}}}